\documentclass[review,authoryear]{elsarticle}

\usepackage{lineno,hyperref,todo,array}
\modulolinenumbers[5]

\journal{Advances in Space Research}

\biboptions{square}

\hypersetup{
     colorlinks = true,
     linkcolor = blue,
     anchorcolor = blue,
     citecolor = blue,
     filecolor = blue,
     urlcolor = blue
     }
\usepackage{verbatim} % to comment out large sections
\usepackage{mathtools} % align equations

 \usepackage[]{pdfcomment} % \pdfcomment{mycomment to insert} 

\usepackage{xcolor} 
\usepackage{float} % use with H to keep tables in its own section

\makeatletter
{\catcode`\`=13
\xdef\@verbatim{\unexpanded\expandafter{\@verbatim}\chardef\noexpand`=18 }
}
\makeatother

\begin{document}

\begin{frontmatter}

% Does not compile as downloaded problem is in the \title
\title{GipsyX/RTGx, A New Tool Set for Space Geodetic Operations and Research \footnote{\copyright{ 2020 This manuscript version is made available under the CC-BY-NC-ND 4.0, \href{http://creativecommons.org/licenses/by-nc-nd/4.0/}{http://creativecommons.org/licenses/by-nc-nd/4.0}}, DOI: 10.1016/j.asr.2020.04.015}}

%% No white space allowed between multiple addresses, group authors per affiliation:
\author[addressJPL]{Willy Bertiger\corref{correspondingauthor}}
\ead{william.bertiger@jpl.nasa.gov}
\author[addressJPL]{Yoaz Bar-Sever}
\author[addressJPL]{Angie Dorsey}
\author[addressJPL]{Bruce Haines}
\author[addressJPL]{Nate Harvey}
\author[addressJPL]{Dan Hemberger}
\author[addressJPL]{Michael Heflin}
\author[addressJPL]{Wenwen Lu}
\author[addressJPL]{Mark Miller}
\author[addressJPL]{Angelyn W. Moore}
\author[addressJPL]{Dave Murphy}
\author[addressJPL]{Paul Ries}
\author[addressJPL]{Larry Romans}
\author[addressJPL]{Aurore Sibois}
\author[addressJPL]{Ant Sibthorpe}
\author[addressJPL]{Bela Szilagyi}
\author[addressJPL]{Michele Vallisneri}
\author[addressIGN,addressIPGP]{Pascal Willis} % no white space allowed between multiple addresses

\address[addressJPL]{Jet Propulsion Laboratory, California Institute of Technology, 4800 Oak Grove Drive, Pasadena CA 91109}
\address[addressIGN]{Universit\'e de Paris, Institut de physique du globe de Paris, CNRS, IGN, F-75005 Paris, France}
\address[addressIPGP]{ENSG-G\'eomatique, IGN, F-77455 Marne-la-Vall\'ee, France}
\cortext[correspondingauthor]{Corresponding author}

% Willy added tableofcontents to make it quick to jump to sections comment out before submitting
\tableofcontents

\begin{abstract}

%{\color{red} NOTE: if you see a blue box in your pdf preview window, try opening the pdf in Adobe Acrobat instead. 
%It is probably because your viewer does not support pdf comments using the pdfcomment package. }

GipsyX/RTGx is the Jet Propulsion Laboratory's (JPL) next generation software package for positioning, navigation, timing, and Earth science using 
measurements from three geodetic techniques:  Global Navigation Satellite Systems (GNSS), Satellite Laser Ranging (SLR), and Doppler Orbitography and Radiopositioning Integrated by Satellite (DORIS); with Very Long Baseline Interferometry (VLBI) under development.  The software facilitates
combined estimation of geodetic and geophysical parameters using a Kalman filter approach on real or simulated data in both post-processing and in real-time. The estimated parameters include station coordinates and velocities, satellite orbits and clocks, Earth orientation, ionospheric and tropospheric delays. The software is also capable of full 
realization of a dynamic terrestrial reference through analysis and combination of time series of ground station coordinates.

Applying lessons learned from its predecessors, GIPSY-OASIS and Real Time GIPSY (RTG),  GipsyX/RTGx was re-designed from the ground up to offer 
improved precision, accuracy, usability, and operational flexibility. We present some key aspects of its new architecture, and describe some of its 
major applications, including Real-time orbit determination and ephemeris predictions in the U.S. Air Force Next Generation GPS Operational Control 
Segment (OCX), as well as in JPL's Global Differential  GPS (GDGPS) System, supporting User Range Error (URE) of $<$ 5 cm RMS; precision 
post-processing GNSS orbit determination, including JPL's contributions to the International GNSS Service (IGS) with URE in the 2 cm RMS range; Precise 
point positioning (PPP) with ambiguity resolution, both statically and kinematically, for geodetic applications with 2 mm horizontal, and 6.5 mm vertical 
repeatability for static positioning; Operational orbit and clock determination for Low Earth Orbiting (LEO) satellites, such as NASA's Gravity Recovery and 
Climate Experiment (GRACE) mission with GRACE relative clock alignment at the 20 ps level; calibration of radio occultation data from LEO satellites for weather forecasting and climate studies; Satellite 
Laser Ranging (SLR) to GNSS and LEO satellites, DORIS-based and multi-technique orbit determination for LEO; production of terrestrial 
reference frames and Earth rotation parameters in support of JPL's contribution to the International Terrestrial Reference Frame (ITRF).

\end{abstract}

\begin{keyword}
\texttt{Multi-technique space geodesy}\sep orbit determination and satellite clocks\sep Precise Point Positioning (PPP)\sep GNSS \sep SLR \sep DORIS
\MSC[2010] 00-01\sep  99-00
\end{keyword}
\end{frontmatter}

%\linenumbers

\section{Introduction}
% willy deleted most of the samples for equations, figures,... see jsar.tex for examples when needed

Over the past thirty years, Global Navigation Satellite Systems (GNSS), starting with GPS, have become ubiquitous in our daily lives. A growing number of scientific, industrial, and public safety applications, spanning  geodesy, geophysics, seismology, atmospheric sciences, weather forecast and climatology, time and frequency transfer,  have come to depend on precision modeling of all aspects of GNSS at the cm and mm accuracy. Very few software packages exist with the capabilities to obtain these highest accuracy results, keeping up with the latest GNSS models and data processing techniques. Fewer still are capable of also modeling and processing data from other space geodetic techniques such as SLR (Satellite laser ranging), DORIS (Doppler orbit determination and radio-positioning integrated by satellite),
or VLBI (Very long baseline interferometry). Recently, JPL has embarked upon a major overhaul of its widely-used space geodetic data analysis software packages, GIPSY-OASIS and RTG, to derive GipsyX/RTGx, with many advanced new capabilities and quality features. This paper is intended for the users and developers of space geodesy tools.

We present the design and some implementation notes for a new, but already widely used software set, and describe its performance in a broad range of 
applications spanning orbit determination and positioning operations, geodesy, and remote sensing. We first give some historical context, which influences any 
large software and research project design. We then present the overall design of the software, and sample use cases with tests of accuracy and precision.

The Jet Propulsion Laboratory (JPL) has a long history in space geodesy and precision orbit determination (POD) for Earth orbiting satellites, and 
GipsyX/RTGx is the 4th major redesign of JPL's GNSS data analysis software. For an overview of POD and measurement processing see the books \citep{Tapley2004} and \citep{ref:bierman1977}. Born from Very Long Baseline Interferometry (VLBI) data analysis 
expertise and software \citep{Sovers1987Masterfit}, the first JPL GPS data models were coded during 1984-1985 in the GPSOMC FORTRAN software 
\citep{Sovers1990GPSOMC}. Immediately thereafter we began development of a more comprehensive software set, the GPS Inferred Positioning 
System and Orbit Analysis Simulation Software package (GIPSY-OASIS, forever nicknamed GIPSY) \citep{Wu1986GIPSY}, also coded in FORTRAN, 
to support the early campaign-style GPS orbit determination (GiG'91 for example \citep{melbourne1993first}) and positioning experiments \citep{blewitt1989carrier}. That very early version of GIPSY 
should be considered a prototype for the more formally designed, enhanced, and documented version \citep{Wu1990GIPSY}, written to support the first 
precision GPS flight experiment on the TOPEX/Poseidon altimetry mission \citep{bertiger1994gps} as well as the continuous GPS orbit determination 
operations at JPL that are the cornerstone for all precision GIPSY applications. %add some more of the early GIPSY references

With the maturation of the GPS constellation, which became fully operational in 1994, real-time operational applications, such as differential 
positioning, started to emerge. The state-space formulation of real-time differential systems, today a universally accepted practice also known as wide 
area differential GPS (WADGPS) or global differential GPS (GDGPS), was conceived and patented at JPL \citep{Yunck1998WADGPS}, and led to the 
creation of the Real Time GIPSY (RTG) software. RTG ported most of GIPSY's precise satellite and signal models from FORTRAN into C, but 
redesigned the software architecture, and in particular the estimation filter, for efficient real-time operations. Successful early demonstration of RTG  
\citep{Whitehead1998satloc} has led the  United States Federal Aviation Authority's (FAA) to adopt JPL's state-space algorithms for its Wide Area 
Augmentation System (WAAS), with RTG as the prototype software for this critical aviation infrastructure, which has since been adopted by several 
other countries as well \citep{bertiger1997real}. % add more WAAS
In 2000, RTG became core software for JPL's GDGPS system, as well as for other commercial systems.

As with any JPL-developed software, GIPSY and RTG are owned by the California Institute of Technology (Caltech). Licensing requests are available from  [\href{https://gipsy-oasis.jpl.nasa.gov/ }{https:// gipsy-oasis.jpl.nasa.gov}]. With hundreds of issued licenses 
to academia (which are free of license fees), industry, and government, GIPSY and RTG are among JPL and NASA's most licensed software, and the recipient of multiple industry 
and NASA awards, including the 2000 NASA Software of the Year Award for RTG, and 2004 Space Technology Hall of Fame induction for GIPSY and 
RTG (as ``Precision GPS Software System" ) [\href{https://spinoff.nasa.gov/Spinoff2008/award_winners.html}{https://spinoff.nasa.gov/Spinoff2008/
award\_winners.html}].

In 2005 Raytheon (as lead), ITT (now L3Harris Corporation), and JPL teamed up to develop an innovative navigation concept of operation for the U.S. Air 
Force next generation GPS operational control segment, known as OCX. The OCX project is a complete overhaul of the GPS operational control 
segment, including new architecture, infrastructure, hardware, and software required to comply with a set of demanding performance and quality 
specifications. The team's proposed navigation concept was patterned after JPL's GDGPS System and built upon JPL's proven GIPSY and RTG 
navigation software technology. In 2010 the team was competitively awarded the \$1B+ project to develop and deploy OCX [\href{http://
insidegnss.com/raytheon-wins-1-5-billion-gps-ocx-contract/}{http://insidegnss.com/raytheon-wins-1-5-billion-gps-ocx-contract/}], \citep{bertigerIONocx2010},  and we set out to 
create JPL's 4th generation GPS data processing software, entitled RTGx. Later, the NASA Space Geodesy Project (SGP) provided additional support 
to enhance post-processing for space geodesy \citep{Merkowitz2018}. All programs synergistically benefited. 

This was an opportunity to modernize the entire geodetic data processing software set at JPL and unify the agile real-time processing of RTG with the 
post-processing flexibility of GIPSY, all within a new architecture that can support all of JPL's, NASA's and the Air Force's OCX current and anticipated 
needs. This unified multi-capability software set was named GipsyX/RTGx (or RTGx/GipsyX, depending on the main use case. The choice of 
capitalization is intentional).

The development of GipsyX/RTGx was mostly completed by 2016.  By 2014 it had replaced RTG as the orbit determination engine for most GDGPS real-time 
GNSS operations, and by 2017 it had replaced GIPSY as the orbit determination engine of all post-processed GPS orbit determination operations at JPL, 
including the reprocessing of the entire record of GPS ground tracking data back to the early 1990s (in order to establish a consistent long-term set of 
orbit and clock solutions in ITRF2014). However, with the ever-changing GNSS landscape and the on-going quest to combine the various geodetic 
measurement techniques, GipsyX/RTGx is constantly being updated and upgraded.

GipsyX/RTGx, as was GIPSY-OASIS, is a member of a small `club' of available precision space geodetic data analysis packages, with diverse capabilities, 
and with a track record of contributing high quality GNSS data analysis products. These include Bernese \citep{berneseV5.2},  EPOS.P8 \citep{Uhlemann2015GFZ}, GAMIT \citep{HerringGAMIT},  GEODYN \citep{pavlisGeodyn2017}, GINS \citep{GINSref}, NAPEOS \citep{springer2011napeos}, and PANDA \citep{shi2010introduction}. %CODE \citep{dach2009gnss},

Since 2018, GIPSY and RTG are no longer supported by JPL. The large GIPSY licensee community is transitioning to GipsyX/RTGx with the help 
of JPL-provided classes, online documentation, tutorials and licensing information on (\href{https://gipsy-oasis.jpl.nasa.gov}
{https://gipsy-oasis.jpl.nasa.gov}) with over 230 licenses of which 80\% are free academic licenses.

\section{Software Design, Overview}

GIPSY's core computational engines were written in FORTRAN with glue code and automation in C-shell, Bourne-Shell, and Perl. RTG was written in 
C. Both GIPSY and RTG implemented precise sets of models consistent with the 2010 Conventions of the International Earth Rotation Services (IERS) 
\citep{petit2010iers}. RTG was designed for lean real-time processing, and did not have post-processing capabilities required by many of NASA's 
science applications and by the geodetic user community. While FORTRAN has been modernized since our original GIPSY writing, it is not currently 
as widely used as C or C++, and is hardly taught at universities. The U.S. Air Force contract for OCX forbade outright the use of FORTRAN.  C does not 
support object-oriented programing, which we deemed necessary for ease of development and maintenance. We chose, therefore, to write  the main 
computational parts of the new software set in C++. We wanted to avoid the hodgepodge of scripting languages used in GIPSY and chose Python3, which has 
a large user base and community support in the sciences and engineering, as the sole scripting language. For small problems and identical use cases (for instance 
precise point positioning with GPS or low Earth orbit determination with GPS) which are implemented in GIPSY and GipsyX/RTG, run times are nearly identical. Larger
problems, that can be handled by GIPSY and GipsyX/RTGx can be faster with the new software, due to optimization and use of parallel computation. 
Table \ref{tbl:CPUtime}, shows almost a factor of three increase in performance for a moderate size problem, the problem is too small for realizing large increases in 
speed with multiple cores/CPUs.
In addition to new use cases, which
are not implemented in GIPSY, there are very large problems that can only be handled in GipsyX/RTGx, simultaneous adjustment of a large gravity field, GNSS orbits, GRACE orbits and 
additional low Earth orbiters. For use cases that are common to GIPSY and GipsyX/RTGx, we expect almost identical precision and accuracy, but a much 
improved user interface. It is the preservation of the accuracy and precision, along with the additional use cases that adds new value to space geodesy.

\begin{table}[htp]
\caption{CPU Time comparison, GIPSY verses GipsyX/RTGx for a moderate size problem, 45 stations, 32 GPS satellites with bias fixing. 
Unloaded Intel\textsuperscript{\textregistered} Xeon\textsuperscript{\textregistered} Gold 6130 CPU at 2.1 GHz. GIPSY does not support 
multiple cores.
}
\begin{center}
\begin{tabular}{c c}
 & CPU Time(minutes) \\ \hline
GIPSY & 45 \\
GipsyX/RTGx 1-core&  16\\
GipsyX/RTGx 2-core&  11\\
\end{tabular}
\end{center}
\label{tbl:CPUtime}
\end{table}

Among the key new capabilities of GipsyX/RTGx are:
\begin{itemize}
\item Complete support for most Linux distributions and Mac OS X; tailored installation on embedded flight hardware
\item Full GNSS satellite and signal modeling, including GPS, GLONASS, BeiDou, Galileo, and QZSS
\item Seamless processing of data from shared memory or files supports both real-time and post-processing operations
%\pdfcomment{Need to be explicit that shared memory, and therefore probably real-time, is not available in the GipsyX release version. Maybe not, since it is available for a license fee}
\item High fidelity data simulator capable of simulating all the data types it can process
\item Efficient square root information filter (SRIF) formulation featuring multi-threading and parallelization with Message Passing Interface (MPI)
\item Hot start capability with archived filter state and covariance
\item Integer GNSS phase ambiguity resolution in post-processing and in real-time 
\item Inertial measurement modeling to optimally combine GNSS and inertial data in real-time and in post-processing
%\item Satellite laser ranging (SLR) data processing
%\pdfcomment{Is this a new capability of GipsyX/RTGx?, No taking it out but leaving comment for Yoaz}
\item Hyperlinked documentation using \href{http://www.doxygen.nl/}{Doxygen} and \href{http://www.sphinx-doc.org/en/master/}{Sphinx} 
\item User input interface featuring inheritance
\item Reference frame determination and time series analysis of Earth station positions
%\pdfcomment{Is this a new capability of GipsyX/RTGx? Reference frame is new}
\end{itemize}

Table \ref{tbl:srcCount} shows the size of the entire GipsyX/RTGx distribution in terms of the number of lines of code. Raw lines are just the total 
number of lines in all the source files. Physical lines are lines that are not comments or blank. The majority of the C++ code occurs in the single 
executable module, ``rtgx'', which models the GNSS data and other tracking data types, and estimates parameters such as station and satellite 
position. We describe the structure of ``rtgx''  below. There are approximately 140  individual executable programs in GipsyX/RTGx. 
Included are many small utilities, such as time format manipulation, time series and data manipulation, trend analysis, data archive access and plotting 
utilities.
Most users will 
only use a small set of these. In addition to the tracking data analysis tools, there is a set of Python3 executables dedicated to time series analysis of 
positions on the Earth and reference frame determination as well as many other small utility tools (for example, to convert between file formats or 
timing formats).

% ran slic took the total at the end
\begin{table}[htp]
\caption{GispyX source lines, 1k=1,000}
\begin{center}
\begin{tabular}{c c c c}
Language & Physical Lines & Raw Lines & Comment Lines \\ \hline
C++ & 202.7k & 353.7k & 103.3k \\
Python3 & 35.6k & 60.4k & 14.6k \\
\end{tabular}
\end{center}
\label{tbl:srcCount}
\end{table}%

We have attempted to apply in this development the many lessons learned from the previous iterations of GIPSY and RTG. GIPSY was written by 
many people under a fairly loose set of coding rules. The main computational pieces of GIPSY were divided into several independent FORTRAN 
modules (orbit integration, signal/Earth model, filter, smoother...). There was considerable inconsistency in the input to these pieces, which were 
mostly FORTRAN namelists. Common information across these FORTRAN executables was not consistently named or configured. To mitigate the 
complex  configuration of the various GIPSY modules, we wrote a Perl script (gd2p.pl) to hide the complexity and act as the interface for most of our external users 
who were doing precise point positioning (PPP) or precise orbit determination (POD) of low Earth Orbiters using GPS \citep{bertiger2010single}. 
GipsyX/RTGx improves upon GIPSY in two ways here. One, there is a uniform input format, referred to as a `tree' input defined in detail below, which has a general structure 
similar to \href{https://yaml.org}{YAML}. We did not use YAML because, at the time of our development, it did not have some object-oriented inheritance 
properties or arbitrary script execution that would allow human readable, compact input for GNSS constellations where many of the satellites have 
identical properties. Second, instead of writing many individual executable modules as in GIPSY, the object-oriented C++ allowed us to write a single 
main executable, rtgx.

For version control, we are using \href{https://subversion.apache.org/}{svn, subversion} \citep{pilato2008version} chosen before distributed version control systems such as git \citep{Chacon2014} were more popular. We believe that documentation is best maintained inside the 
source code; thus, we chose to use \href{http://www.doxygen.nl/}{Doxygen} for C++ code, and \href{http://www.sphinx-doc.org/en/master/}{Sphinx} for 
Python. Doxygen and Sphinx both output html documentation and it is easy to link the two sets of documentation. The mathematical description for the 
radio-metric signals, solid tides, Earth orbiting satellite forces, ordinary differential equation integrator, etc., are best written in \LaTeX. Our html 
documentation includes a pdf  containing the mathematical description, generated with \LaTeX, with links to the source code documentation.

Having a single main executable, nearly all use cases can be constructed as a very thin wrapper around a single command line:
\begin{verbatim}
rtgx myInput.tree
\end{verbatim}

To automate the most common tasks carried out by our user community, namely precise static point positioning with single receiver ambiguity resolution, we have provided a Python3 wrapper, where the only required input is a RINEX2 (Receiver Independent Exchange Format) or \href{ftp://igs.org/pub/data/format/rinex303.pdf}{RINEX3 GNSS data file}. 
\begin{verbatim}
gd2e.py -rnx my.rnx
\end{verbatim}
This is the analog of GIPSY's gd2p.pl automation script, and it contains many enhancements over its predecessor in terms of ease of use and documentation.

\section{User interface - the input tree}

Several of our executables, especially those requiring more complicated and finely controllable user inputs, 
are configured by a single interface called a tree. A tree is a text file with a hierarchical structure identified 
using Python-like indentations that consist of roots (the highest level), branches, and leaves (the lowest level; 
the leaves of the tree are the parts of a tree with no branches and are generally the ones that contain specific 
data). See \href{file:ElectronicS1_UserInterface.pdf}{electronic supplement 1} for details.

\section{Main C++ Software Modules/Classes}

The computationally intensive software breaks into several broad categories:
\begin{itemize}
\item Data editing
\item Orbit Integration and force models
\item Signal Models
\item Earth Models
\item Filter, optimized fit to the linearized model forward in time
\item Smoother, optimized fit to the linearized model over all time
\end{itemize}
All of these items are implemented as C++ classes or functions in a single executable module, rtgx, except for data editing, which is provided by the ``gde'' module.

\subsection{gde Data Editor}

GNSS data must be edited for phase breaks and gross outliers before it is processed to fit model parameters. The GNSS Data Editor (gde), is a C++ 
code that implements two distinct algorithms: turbo-edit \citep{blewitt1990automatic}, and simple continuity checks of linear combinations of phase data 
based on low-degree polynominal fits.  Turbo-edit looks at averages of $pseudorange-phase$  measurements to detect jumps in the phase 
measurements. It is well suited to low rate data, such as the large number of RINEX files typically recorded at a 30-sec rate.  When data rates are 
high, for example sampled every second as often seen in real-time applications, it is possible to detect phase breaks simply by monitoring the change 
in time of differences in phase observations on two frequencies.  Such a difference cancels out receiver and transmitter clock jumps. Removing a low 
order time polynomial from these differences effectively removes the slow-changing  ionospheric signal in this difference. Single differences of dual-
frequency combinations across two satellites are also available as an option. 

\subsection{Orbit Integration and force models} \label{sec:oi}

To model the path of a satellite in orbit about the Earth, we implement precise force models that include gravitational forces and 
surface forces on the satellite. 
These forces may contain model parameters. Given initial values for the satellite epoch state (position and velocity at the initial 
epoch time) and parameter values, the second order differential system of equations for the time evolution of the satellite state 
and the partial derivatives of the satellite epoch state with respect to the model parameters are integrated numerically forward in 
time.
Force models represented in GipsyX/RTGx include the Earth's gravity field as a spherical harmonic expansion to arbitrary degree 
and order including time variability of these coefficients, gravitational effects of Earth solid tides, pole tides and ocean tides, first 
order General Relativistic acceleration and atmospheric drag (DTM-2000, \citep{bruinsma2003dtm}).  Conical models for solar eclipse 
by the Earth and Moon are implemented. Generic Earth albedo (\citep{ref:knocke89, ref:knocke&ries&tapley88}) and solar radiation 
pressure (\citep{milani1987non}) models are available through a flexible tree input interface which enables the specification of arbitrary 
sets of panels along with their orientation and reflective properties. More specific satellite radiation pressure models, including some 
empirical GPS solar radiation pressure models specially developed at JPL  (\citep{bar2005new} and \citep{sibthorpe2011evaluation}), 
may also be input either as arbitrary Fourier series, or bi-linearly interpolated tables dependent on Sun azimuth and elevation angles in
the satellite body-fixed frame.

In order to accurately compute surface force models, spacecraft orientation is required, and may be input in a variety of ways, including 
time-series of attitude quaternions and, especially useful for simulations, dynamic pointing specified via mathematical operations on vectors from the 
satellite to the Sun and Earth, and the satellite's velocity with a defined input calculus. Attitude may also be selected from a number of specific models including
those for GPS (\citep{bar1996new, bar1996fixing}), GLONASS (\citep{dilssner2011GLONASS}), Galileo (\href{https://www.gsc-europa.eu/support-to-developers/galileo-iov-satellite-metadata}{https://www.gsc-europa.eu/support-to-developers/galileo-iov-satellite-metadata}) and 
configurable yaw steering and orbit normal modes for BeiDou geostationary and inclined geostationary satellites (\citep{bdsAtt2017}). For other orbiting satellites including altimetry missions, 
attitude may be selected from models or quaternions. 

We have implemented a general-purpose ordinary differential equation (ODE) solver which we use to obtain
numerical solutions of the time evolution of the satellite state.  This ODE solver includes 
multistep methods \citep{ref:Hairer1993} as well as high-order embedded Runge-Kutta methods \citep{ref:Prince1981}.
The solver employs typical local-truncation-error-estimate adaptive step size techniques.
In order to allow for discontinuities in the integrated source functions (e.g., the solar radiation pressure transition between shadow regimes,
 and satellite thruster firings have discontinuous time derivatives) and yet still
employ high-order integration techniques, the ODE solver employs a model-specific discontinuity detection logic, following the``gstop-function'' approach specialized at JPL by Fred Krogh for integrating spacecraft trajectories (\citep{Krogh1972}) \href{https://mathalacarte.com/fkrogh/pub/dxrkmeth.pdf}{https://mathalacarte.com/fkrogh/pub/dxrkmeth.pdf}. 

\subsection{Signal Model}

The basic signal model is the time of propagation from a transmitter to a receiver. For GNSS range measurements 
recorded at receiver time, $\tilde{t_r}$, determined by the receiver's clock and transmitted at time $\tilde{t_t}$, we 
model the measured range as 
\begin{equation} \label{eq:sigModel}
R = c(\tilde{t_r} - \tilde{t_t}) + d_{trop} + d_{iono}
\end{equation}
where $c$ is the speed of light, $d_{trop}$ is a delay due to troposphere and $d_{iono}$ is the delay due to the ionosphere. For many GNSS 
measurements the first order delay of the ionosphere is removed by a dual-frequency combination of the data \citep{parkinson1996global} 
and a model for higher order effects \citep{kedar2003effect}. The troposphere is modeled as a delay at zenith plus 
gradient parameters (\citep{ref:bar1998estimating},  \citep{bohm2004vienna}, \citep{bohm2006global}, \citep{bohm2015gpt2w}). The difference in 
the time of reception, $\tilde{t_r}$, and the time of transmission, $\tilde{t_t}$, in equation \ref{eq:sigModel} can be 
modeled as

\begin{equation} \label{eq:tDiff}
\begin{aligned}
\tilde{t_r} - \tilde{t_t} &=  \tilde{t_r} - \bar{t_r} \\
                                     &+ \bar{t_r} - t_r \\
                                     &+ t_r - t_t \\
                                     &+ t_t - \bar{t_t} \\
                                     &+ \bar{t_t} - \tilde{t_t}
\end{aligned}
\end{equation}
where $\tilde{t_r} - \bar{t_r}$ is the difference between the time on the receiver clock and proper time(time clock would 
have read with no errors) and $\bar{t_r} - t_r$ is the difference between proper time and coordinate time (General 
Relativistic effects,  \citep{Moyer1971}, \citep{ref:thomas}, \citep{Moyer_1981}). The last two lines of eq. \ref{eq:tDiff} contain similar differences for the 
transmitter. The middle term, difference in coordinate time, $t_r - t_t$, is modeled as the geodesic distance (General 
Relativity with the only mass being the Earth's point mass) between the transmitter and receiver phase centers. The fit 
to measurements of the range is optimized by linearizing about the nominal parameter values in equations 
\ref{eq:sigModel}, \ref{eq:tDiff}, first analytically taking partial derivatives with respect to the parameters and then in the 
case of satellites using the numerically calculated partials in sec. \ref{sec:oi}.

For multi-GNSS, eq. \ref{eq:sigModel}, must be modified due to different delays of ranging codes through the receiver for 
different constellations. The code adds a bias parameter for each receiver by constellation \citep{odijk2013characterization}. One constellation must be 
held as reference to prevent singularity. Typically we hold GPS as reference, but it is arbitrary.

The model for GNSS phase data is identical to range, except two terms must be added to eq. \ref{eq:sigModel}
\begin{equation} \label{eq:sigModelPhase}
\phi = c(\tilde{t_r} - \tilde{t_t}) + d_{trop} + d_{iono} + \omega(t_r) + B^t_r
\end{equation}
where $\phi$ is the modeled phase, $B^t_r$, is an arbitrary phase bias between the transmitter and receiver over time periods where the receiver has not lost lock, and $\omega(t_r)$ is the phase windup \citep{wu1993effects}. The $B^t_r$ may further be modeled as integer number of wavelengths and fractional hardware delays in the given receiver and transmitter; see \citep{ref:blewitt1989carrier} for treatment of network ambiguity resolution and \citep{bertiger2010single} for treatment of the integer ambiguities with a single receiver.

\subsection{Earth Models}

For receivers or transmitters that are not orbiting the Earth and are either attached to or associated with the Earth's crust, we must 
model the deformation of the Earth's crust and the orientation of the Earth in inertial space since satellite positions are integrated 
in inertial space. GipsyX/RTGx implements the IERS standards \citep{petit2010iers}. Adjustable parameters include the Earth 
orientation parameters, polar motion and hour angle.

\subsection{Filter, Smoother, Ambiguity Resolution}

Except for single receiver use cases, most of the computation time is dominated by filtering, smoothing and ambiguity resolution
functions. If we treated all the parameters as constant in time, the filter and smoother parts of the code would just be a standard least 
squares; but many of the parameters do vary in time, for instance the zenith troposphere delay or the receiver and transmitter 
clock errors. Many force parameters, including empirical accelerations affecting satellite dynamics, are also best treated 
as stochastic variables in a process called reduced-dynamic orbit determination \citep{yunck1990precise, 
wu1991reduced}. We chose to handle these time variations as a first order Markov process with a forward-in-time filter 
implemented as a Square Root Information Filter (SRIF) detailed in Bierman \citep{ref:bierman1977}. The square root 
implementation allows for better conditioned matrices, and the use of Householder (orthogonal transformation in multi-dimensional space) transformations in our 
implementation of the Bierman algorithms lends itself naturally to more efficient use of processor specific pipelining 
\citep{jeong2012performance}, when available, along with the Message Passing Interface (MPI \citep{gropp1999using}) 
for processing across multiple CPU cores and computer clusters (Beowulf \citep{sterling2002beowulf}). 

The forward filter is the best fit of all the past data up to the current epoch. In order to have the parameters fit the 
future as well as the past data optimally, one needs to smooth the information back in time. We again follow Bierman's 
algorithm \citep{ref:bierman1977} for a SRIF smoother, which uses a series of Givens (two-dimensional orthogonal rotation)  and Householder 
transformations. We again optimize efficiency using machine-specific-low-level routines and MPI 
where appropriate. After editing for outliers in the smoother, integer ambiguities may be constrained for 
GNSS data. For single receiver ambiguity resolution, we follow the procedure in \citep{bertiger2010single} extended to 
multi-GNSS, applying constraints to single differences. 
For multiple GNSS receivers, we can resolve double difference integer ambiguities following \citep{blewitt1989carrier} 
extended to multi-GNSS to constrain the double-differenced phase ambiguities, as well as adding single-receiver 
constraints when appropriate external information is available. The single receiver methods are related but not 
identical to methods developed in
 \citep{laurichesse2009integer, laurichesse2011cnes}, \citep{Loyer2012}. Our ambiguity resolution algorithms, do not currently extend to GLONASS
 Frequency Division Multiple Access (FDMA) signals.

\section{Main executable, rtgx}

A detailed description of the main loop of the primary executable is contained in 
\href{file:Electronic2_rtgxMain.pdf}{electronic supplement 2, rtgx main loop}.

\section{Sample Use Cases, Accuracy, Precision}

\subsection{Precision Post-Processed GNSS Orbit and Clock Determination}

\subsubsection{JPL GPS Post-Processed Operational Products}
\label{OpsProducts}

JPL  delivers to the geodetic community, and for combination by the International GNSS Service (IGS) \citep{Johnston2017}, two sets of GPS products characterized by different latencies, precision and accuracy 
levels, but all based on filtering and smoothing of ground tracking data in post-processing.``Rapid" products are generated and distributed daily. ``Final" 
products are generated weekly and typically have a 14-day latency. We also distribute ``Ultra-rapid" products, generated every hour.  While ultra-rapid and rapid products are tightly tied to the IGS 
realization of the ITRF by fixing the coordinates of a subnetwork of stations in the POD process to their ITRF coordinates, Final 
products come in three flavors. Fiducial-free products, identified by the suffix `nf' (non-fiducial) in their names, are not tied to any specific 
terrestrial frame; the positions of the stations constituting the network are left floating with a 1 km a priori sigma, so that the frame of 
these solutions changes from day-to-day. No-net-rotation (`nnr' suffix) products are such that 3 no-net-rotation (relative to the ITRF) constraints are enforced in 
their generation process. Final products with no suffix are tied to the ITRF by means of 7 constraints (3 rotations, 3
translations and 1 scale parameter) and are referred to as NNRTS (no-net-rotation, translation, or scale) or 'fiducial' products, though their
fiducialization differs from Rapid and Ultra-Rapid. Every set of products contain files with consistent orbital state estimates, transmitter clock estimates,
spacecraft attitude information, Earth rotation parameter (ERP) estimates, and widelane phase biases information. In addition, the no-net-rotation 
and fiducial-free solutions are associated with coordinate transformation files referred to as x-files. These x-files contain the set of 7 Helmert parameters (3 small rotations, 3 translations, and 1 scale parameter) 
needed to transform from the frame of the day to the ITRF. The format of these different products is detailed in the GipsyX software 
documentation. 
Following extensive testing and validation, the JPL GNSS analysis center transitioned seamlessly from products generated using 
the GIPSY software package to products created using GipsyX on January 29, 2017. More recently, the entire span of 1994 
through 2018 were reprocessed using GipsyX, consistent with IGS repro2 standards. All products are available at \url{https://sideshow.jpl.nasa.gov/pub/
JPL_GNSS_Products}.

Table \ref{tab:jplProducts} displays the characteristics of each type of post-processed product. Latencies are indicated as well as commonly used 
performance metrics. JPL orbit and clock products cover 30 hours centered at noon, so that each daily solution overlaps with the 
next-day solution by 6 hours. After removing 30-minute tails at both ends of this overlap period, RMS statistics on orbit and clock differences are 
computed in the 5-hour overlap period as an internal measure of the precision of the product. The long-term medians of the RMS clock and orbit differences in the overlap is shown as ``Precision'' in Table \ref{tab:jplProducts}. 
We measure accuracy in Table \ref{tab:jplProducts} by comparing JPL's orbit and clock solutions with the combined final IGS orbit and clock solutions for the same 
days. Although we have listed this as ``Accuracy" in Table \ref{tab:jplProducts}, we note that the IGS Final product is a weighted average of all the contributing analysis centers including JPL.  

\begin{table}[H]
\centering
\begin{tabular}{|>{\centering\arraybackslash}m{1.5cm}|>{\centering\arraybackslash}m{1.8cm}|>{\centering\arraybackslash}m{1.8cm}|>{\centering\arraybackslash}m{1.8cm}|>{\centering\arraybackslash}m{1.8cm}|>{\centering\arraybackslash}m{1.8cm}|}
\hline
\textbf{Product} & \textbf{Latency} & \textbf{Orbit Precision\newline(cm)} & \textbf{Clock Precision\newline(cm)} & \textbf{Orbit Accuracy\newline(cm)} & \textbf{Clock Accuracy\newline(cm)} \\
\hline
Ultra & $<$ 2 hours & 3.4 & 3.0 & 2.2 & 4.3 \\
\hline
Rapid & next day by 16:00 UTC & 2.6 & 2.4 & 2.0 & 4.0 \\
\hline
Final & $<$ 14 days & 2.3 & 2.3 & 1.9 & 3.8 \\
\hline
\end{tabular}
\caption{JPL post-processed operational products. Units for all precision and accuracy metrics are centimeters. Statistics were computed from time 
of switch (Jan. 29, 2017) to Apr. 20, 2019. Each clock metric is the median value of the daily root-mean-
square values of the overlaps/differences across the GPS satellites after removing a linear trend from the entire GPS 
constellation to account for reference clock differences. Each orbit metric is the median value of the daily median of the 3D RSS of 
RMS positions across the GPS constellation.} 
\label{tab:jplProducts}
\end{table}

\begin{figure}[htbp]
\begin{center}
\includegraphics[width=0.8\textwidth]{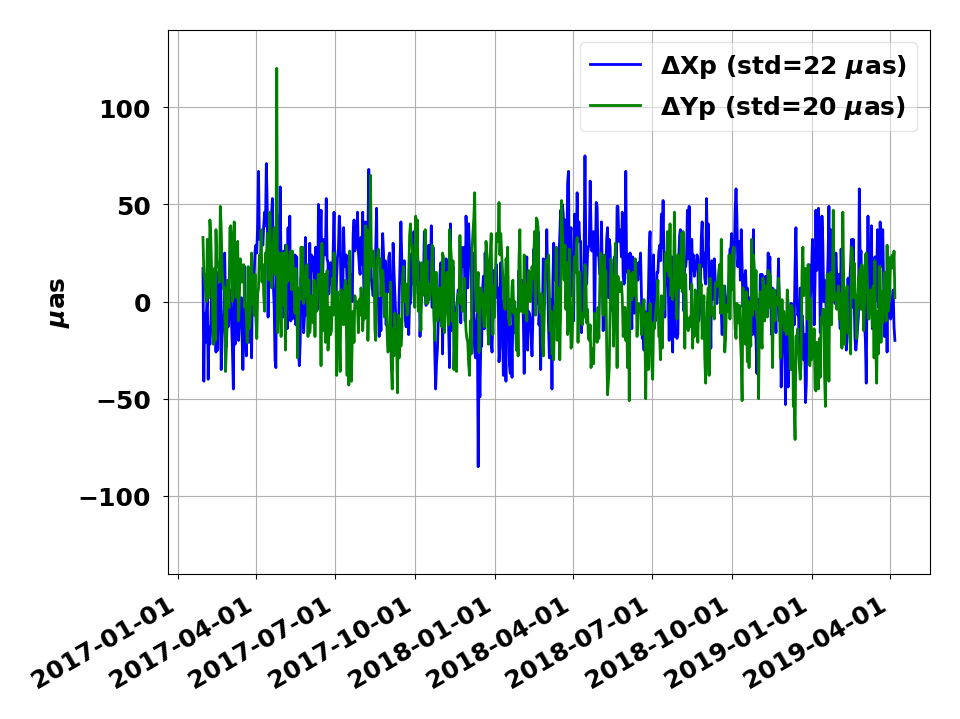} 
\caption{Differences between the Earth's pole coordinates (Xp, Yp) determined by JPL's Final no-net-rotation solutions and values from the IGS final combined solution. The start date of the period covered corresponds to the date of the operational transition from GIPSY to GipsyX at the JPL IGS Analysis Center. The scatter for both coordinates, once converted to equivalent linear distance, is of the order of 0.6 mm at the surface of the Earth.}
\label{fig:pmDeltas}
\end{center}
\end{figure}

\begin{figure}[htbp]
\begin{center}
\includegraphics[width=0.8\textwidth]{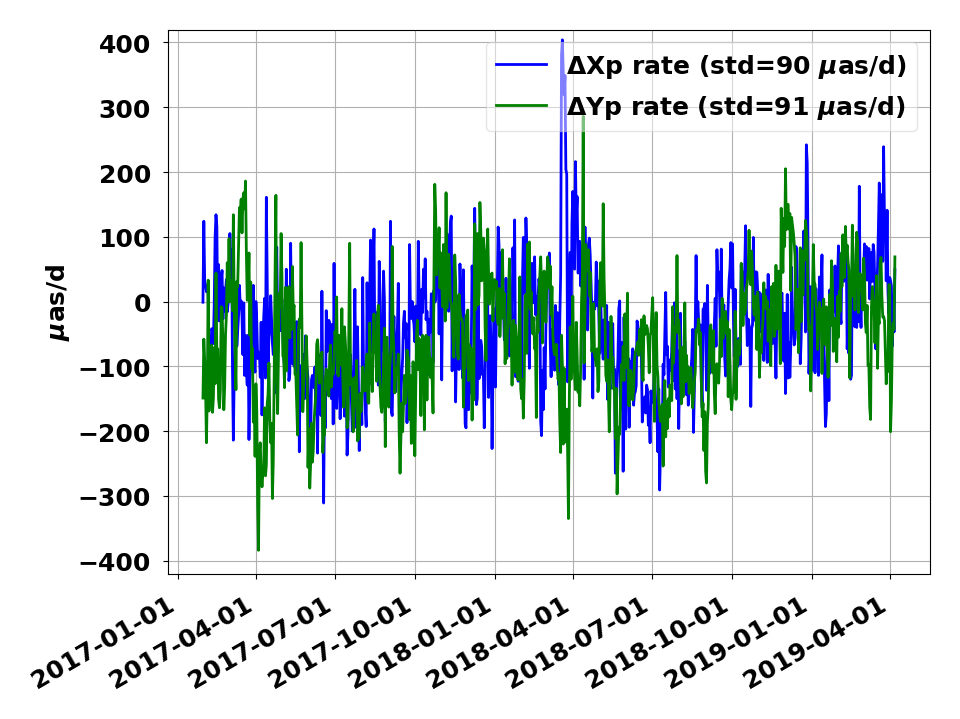}
\caption{Differences between the Earth's pole coordinate (Xp, Yp) rates determined by JPL's Final no-net-rotation solutions and the IGS final combined ERP product. The start date of the period covered corresponds to the date of the operational transition from GIPSY to GipsyX at the JPL IGS Analysis Center. The scatter for both rates correspond to an equivalent linear distance of about 2.8 mm/day at the surface of the Earth.}
\label{fig:pmrDeltas}
\end{center}
\end{figure}

\begin{figure}[htbp]
\begin{center}
\includegraphics[width=0.8\textwidth]{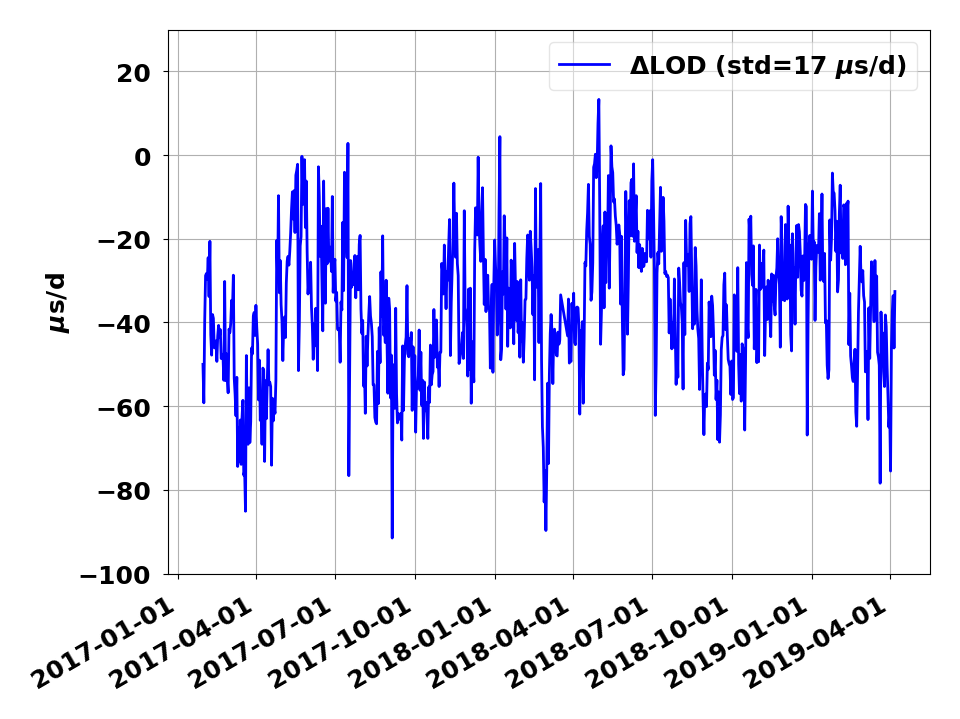}
\caption{Differences between Length-of-Day recovered by JPL's Final no-net-rotation solutions and LOD final combined solution distributed by the IGS. 
The start date of the period covered corresponds to the date of the operational transition from GIPSY to GipsyX at the JPL IGS Analysis Center. 
The scatter around the mean is of the order of 16$\mu$s per day, equivalent to 1.2 mm at the surface of the Earth over a day.
The visible -35$\mu$s/d bias, equivalent to 2.6 mm at the surface of the Earth over a day, is attributed to orbit modeling errors and in particular solar radiation pressure modeling errors in GNSS data processing, as discussed in \citep{sibthorpe2011evaluation} and \citep{ray1996lod} for instance.}
\label{fig:lodDeltas}
\end{center}
\end{figure}

As mentioned above, each set of products includes a file containing estimates of the Earth rotation parameters. The ERPs consist of 
two biases and two rates to describe the motion of the Earth's pole with respect to the Earth's crust: Xp, Yp and their respective rates. The 
excess length-of-day (LOD) is defined as the time derivative of UT1-UTC, where UT1 is the time scale associated to the total rotational 
phase angle of the Earth and UTC is the Universal Time Coordinated. All 6 parameters are directly observable by GNSS, with the exception 
of UT1-UTC. As a result, the polar motion coordinates and rates along with LOD are freely estimated in the GNSS precise orbit 
determination process whereas UT1-UTC is tightly constrained to a nominal value reported in the IERS combined ERP solution, the Bulletin 
A file \citep{iersbulla}. Operational processes at JPL rely on the predicted portion of Bulletin A for a priori values for Earth orientation parameters, while reprocessing campaigns typically use the 
Bulletin A final values of UT1-UTC. All JPL estimates are routinely contributed to the IGS for their combined ERP products. The IGS, in turn, 
contribute these combined GNSS ERP products to the IERS for their final multi-technique combined products. Figures \ref{fig:pmDeltas} 
through \ref{fig:lodDeltas} show the differences between the JPL Final no-net-rotation solutions (contributed to the combined IGS products) 
relative to the final combined IGS solution for Earth Rotation Parameters. The noise in the differences in the polar motion values amounts to 
about 20 $\mu$as for both coordinates, which corresponds to 0.6 mm at the surface of the Earth. The scatter in the differences of the LOD 
estimates is higher at 16 $\mu$s/day, translating into 1.2 mm at the surface of the Earth over one day. These numbers are consistent, even slightly better 
for the pole coordinates, than the statistics reported by \citep{rebischung2016igs}: they found that the weighted RMS of the IGS analysis centers (AC) pole coordinate 
residual time series range between 25 to 40 $\mu$as, while the  weighted RMS of the AC LOD residual time series range from 8 to 20 $\mu$s/day. 
Over the timespan extending from Jan. 29, 2017 to Apr. 20, 2019 (since the transition from JPL's legacy software to GipsyX), the scatter in the differences between the JPL Xp, Yp,
 and LOD solutions relative to 6 other IGS analysis centers using different software packages ranges from 27 to 49 $\mu$as,  26 to 44 $\mu$as, and 13 to 18 $\mu$s/day, respectively.
Again, these statistics are in full agreement with the numbers cited in \citep{rebischung2016igs}.
Orbit modeling differences, e.g., in solar radiation pressure modeling, are known to impact the accuracy of GNSS-based ERP determination. In particular, they could explain the 
bias visible in Fig. \ref{fig:lodDeltas}. The JPL versus IGS ERP differences may be compared directly to the published uncertainties of the final 
combined ERP values from the IERS Bulletin A relative to the IGS combined series \citep{bizouard2017combined}. Table 4 of \citep{bizouard2017combined} 
shows Xp and Yp scatter of IGS combined at 31 and 27 $\mu$as respectively and 10 $\mu$s for LOD for ITRF 2014 from 2010-2015.  

\subsubsection{Four Constellation GNSS Orbits and Clocks}

A greater number of satellites may provide significant benefits to users, including additional coverage in urban canyons, and perhaps better estimates of 
parameters such as troposphere, Earth orientation, and geocenter. To this end, many IGS Multi-GNSS Experiment (MGEX, \citep{montenbruck2017multi}, \url{http://mgex.igs.org/}) analysis centers are now 
routinely processing multiple constellations, particularly GPS, BeiDou, Galileo and GLONASS, although as yet no official combined product is available. It is 
worth considering however, that processing new satellites might also have unintended impacts on, e.g. GPS quality compared to existing GPS-only products, 
because such new satellites may require individual treatment to allow for accurate modeling of their specific observation types (e.g. GLONASS frequency dependent code biases) 
and the spacecraft themselves (e.g. solar radiation pressure, attitude and antenna offsets \citep{montenbruck2015metadata}). Although JPL has been 
processing these constellations for some years now in real-time (see section \ref{RtProducts}), we are still refining our station-data selection procedures
for fully automated post-processed products, as well as our force model choices, alongside internal quality testing, before making an operational version corresponding to one 
or more of our GPS-only product lines (see section \ref{OpsProducts}). Such internal testing showed a dramatic improvement in quality as of GPS week 2056 (starting June 2, 2019) 
with the release of a new IGS ANTEX file (\url{https://kb.igs.org/hc/en-us/articles/216104678-ANTEX-format-description}) which, for the first time, includes antenna 
corrections for all satellites from the four major constellations. With our most recent fully automated software, we have produced daily, Rapid-like 4-constellation products 
for the month of August, 2019, using 120 stations for GPS and 80 for the remaining constellations. Accuracy is more difficult to measure in this case owing to the lack of a 
provably more accurate multi-GNSS product with which to compare, whether from IGS MGEX or elsewhere, and therefore we leave off such an assessment until a future 
date. Individual MGEX analysis center contributions may continue to exhibit significant differences until consistent antenna information, only relatively recently available, 
begins to see widespread use. Our four-constellation products contain an average of 102 satellites per day, approximately comprised of 31 GPS, 28 Beidou, 22 Galileo 
and 21 GLONASS. While this dataset is limited, it gives a preliminary indication of the sort of performance that might be expected from multi-GNSS products processed 
with GipsyX/RTGx, which should also improve with time as models and procedures are further enhanced. 

Table \ref{tab:jplGNSSproducts} displays the overlap precision statistics of each constellation/sub-constellation along with statistics for the JPL GPS Rapid  products described in section \ref{OpsProducts} over the same time period for comparison. As with Table \ref{tab:jplProducts}, each daily solution 
overlaps with the 
next-day solution by 6 hours. After removing 30-minute tails at both ends of this overlap period to avoid edge effects, statistics on orbit and clock differences  are 
computed in the 5-hour overlap period as an internal measure of the precision of the product. The long-term medians of the clock and orbit differences in the 
overlap is shown as ``Precision'' in Table \ref{tab:jplGNSSproducts}. 

\begin{table}[H]
\centering
\begin{tabular}{|>{\centering\arraybackslash}m{2.0cm}|>{\centering\arraybackslash}m{1.8cm}|>{\centering\arraybackslash}m{1.8cm}|>{\centering\arraybackslash}m{1.8cm}|}
\hline
\textbf{Product} & \textbf{Orbit Precision\newline(cm)} & \textbf{Clock Precision\newline(cm)} \\
\hline
GPS (JPL Rapid) & 1.9  & 1.9  \\
\hline
GPS (4-constellation) & 2.0  & 1.5  \\
\hline
BeiDou GEO & 67.9 & 15.3 \\
\hline
BeiDou IGSO & 10.2  & 7.8 \\
\hline
BeiDou MEO & 5.3  & 5.4 \\
\hline
Galileo & 2.8  & 1.4 \\
\hline
GLONASS & 5.5  & 1.1$^*$ \\
\hline
\end{tabular}
\caption{Daily overlaps of JPL four-constellation products (GPS, BeiDou, Galileo and GLONASS). GPS (JPL Rapid), provided
for comparison, is computed using JPL's standard GPS-only Rapid products described in section \ref{OpsProducts}. Units for all precision 
metrics are centimeters. All statistics were computed from Aug. 1 to Aug. 31, 2019. Each clock metric is the median value of the daily 
root-mean-square values of the overlaps/differences across the GNSS satellites after removing a linear trend from the entire GPS 
constellation to account for reference clock differences, a subsequent linear trend from each of Galileo and BeiDou to account for 
constellation bias reference differences, and a linear trend from each GLONASS satellite due to the presence of range biases -- 
$^*$likely making the GLONASS clock overlaps spuriously small. Each orbit metric is the median value of the daily 
median of the 3D RSS of RMS positions across the GNSS constellations.  } 
\label{tab:jplGNSSproducts}
\end{table}

\subsection{Real-time GNSS Orbit and Clock solutions, GNSS Differential Corrections}
\label{RtProducts}

The real-time configuration of GipsyX/RTGx is typically called just RTGx. RTGx has been driving real-time orbit determination 
processes in the GDGPS System since 2014 (RTG was the engine prior to that, with some overlap). These include GPS-only orbit determination 
filters as well as GPS plus various other GNSS constellations. A typical GNSS orbit determination filter may include 1 Hz phase 
and pseudorange measurements from ~150 ground sites tracking 32 GPS satellites, 24 GLONASS satellites, ~35 BDS satellites, 
and ~24 Galileo satellites (as of June 2019). Such a filter produces orbital states every 60 seconds and clock solutions at 1 Hz, with a 
latency(difference in the time corrections are available to a user $-$ epoch of the data used for the clock correction) that never exceeds ~6 seconds relative to the measurement epochs. GNSS accuracy is typically quoted in terms of User Range Error 
(URE), eq. \ref{eq:ure}, where $h$ is the radial orbit error, $c$ is the cross-track error, $l$ is the along-track error, and $clk$ is the clock error:

\begin{equation}
\label{eq:ure}
URE = \sqrt{(h - clk)^2 + (l^2 + c^2)/50}
\end{equation}

\begin{figure}[htbp]
\begin{center}
\includegraphics[width=0.8\textwidth]{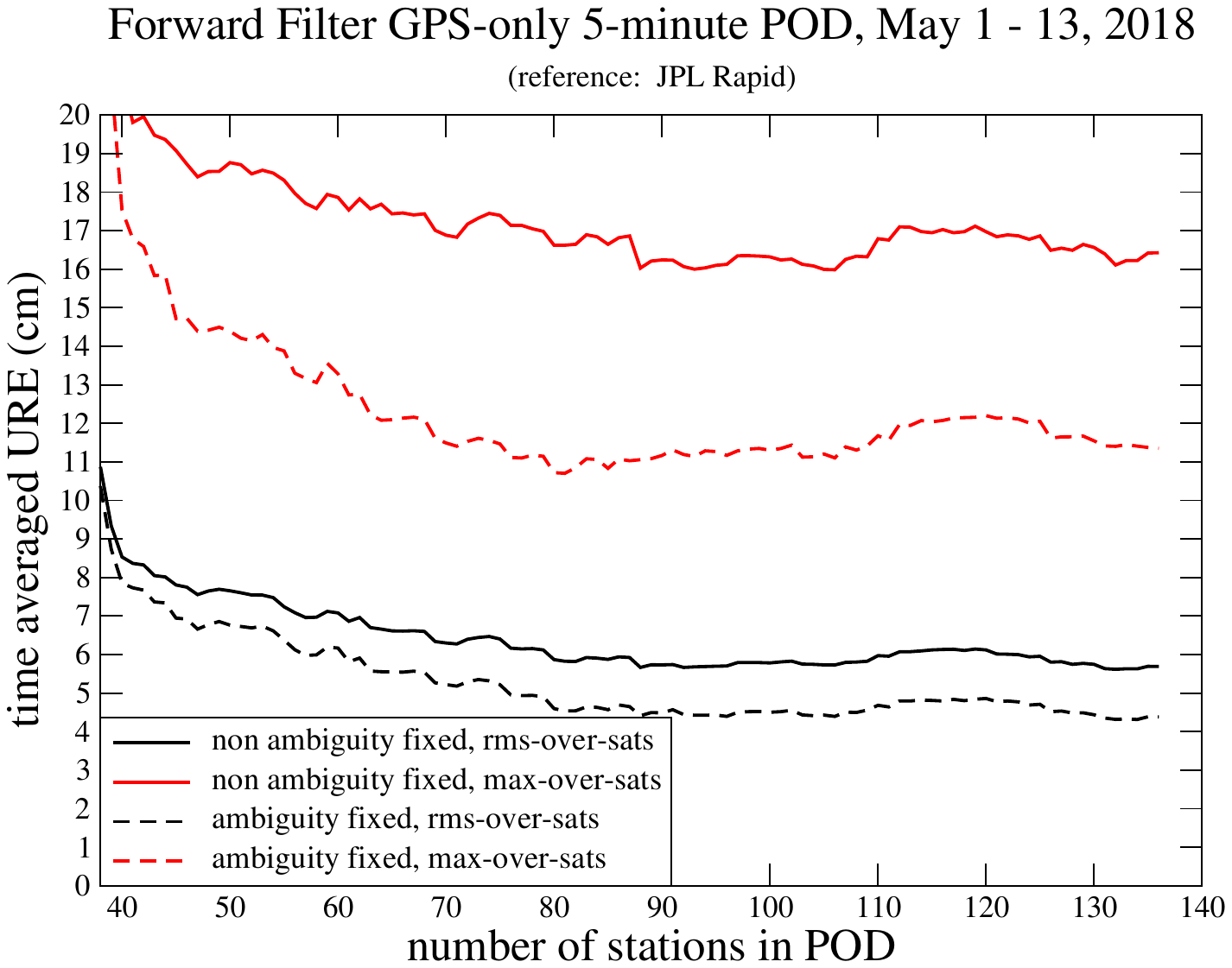} % requires the graphicx package
\caption{RMS URE (black curves) and worst URE (red curves) across all GPS satellites as a function of tracking network size for real-time orbit determination by RTGx within the GDGPS System during May 1 - 13, 2018. Significant improvement due to integer phase ambiguity resolution (dashed curves) is evident relative to the solutions with float ambiguity resolution (solid curves). The reference solutions are provided by JPL's daily post-processed solutions (see Table \ref{tab:jplProducts} above for post-processing accuracy). The RMS URE approaches 5 cm with ambiguity resolution and 90 or more real-time tracking sites.}
\label{fig:GDGPSURE}
\end{center}
\end{figure}

For the full derivation, see the appendix in
 \citep{zumbergeparkinson1996global}.  Here we have used the standard approximations of the coefficients. The real-time URE relative to post-processed solutions is typically ~5 cm RMS (Fig. \ref{fig:GDGPSURE}).

\begin{figure}[htbp]
\begin{center}
\includegraphics[width=0.8\textwidth]{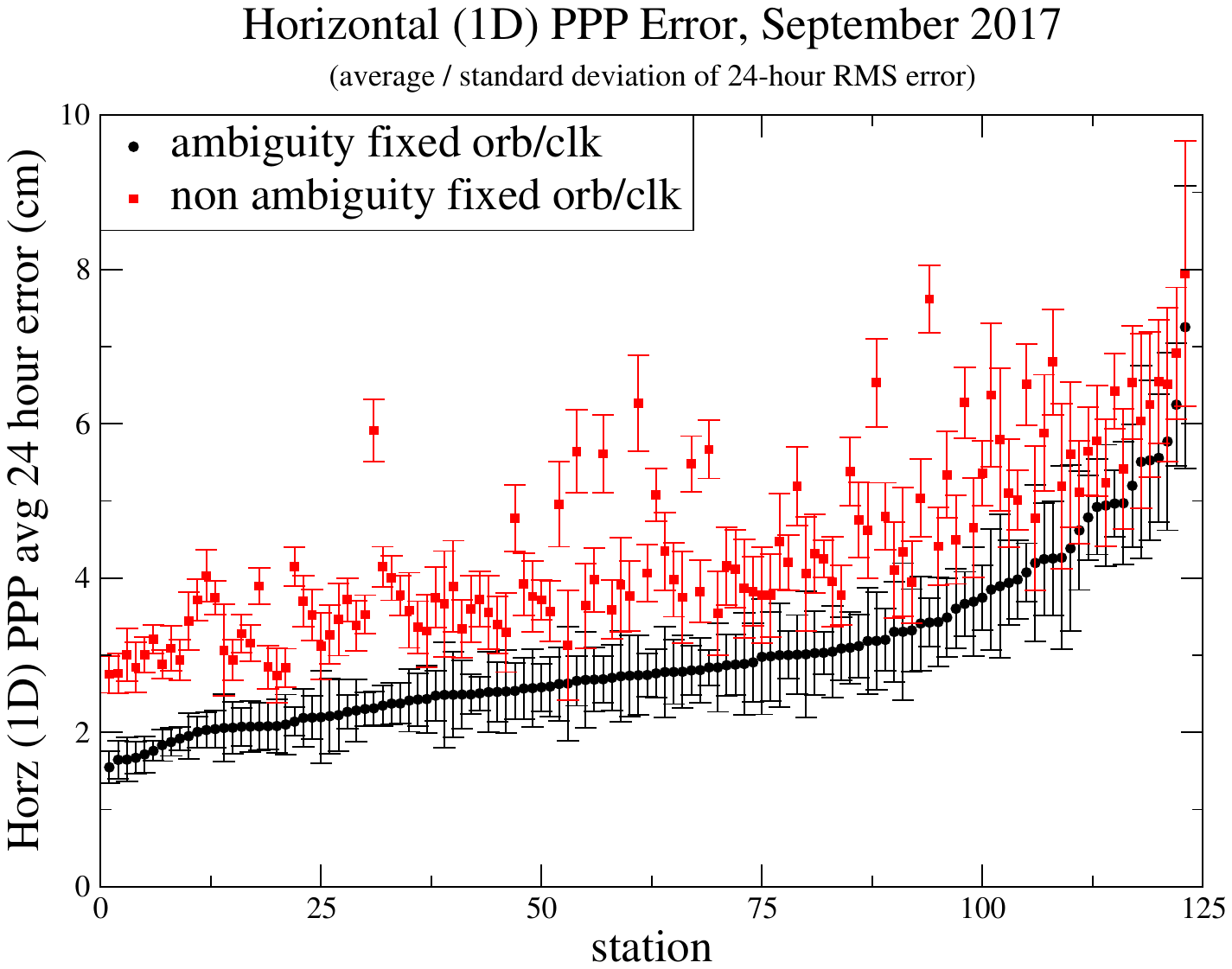} % requires the graphicx package
\includegraphics[width=0.8\textwidth]{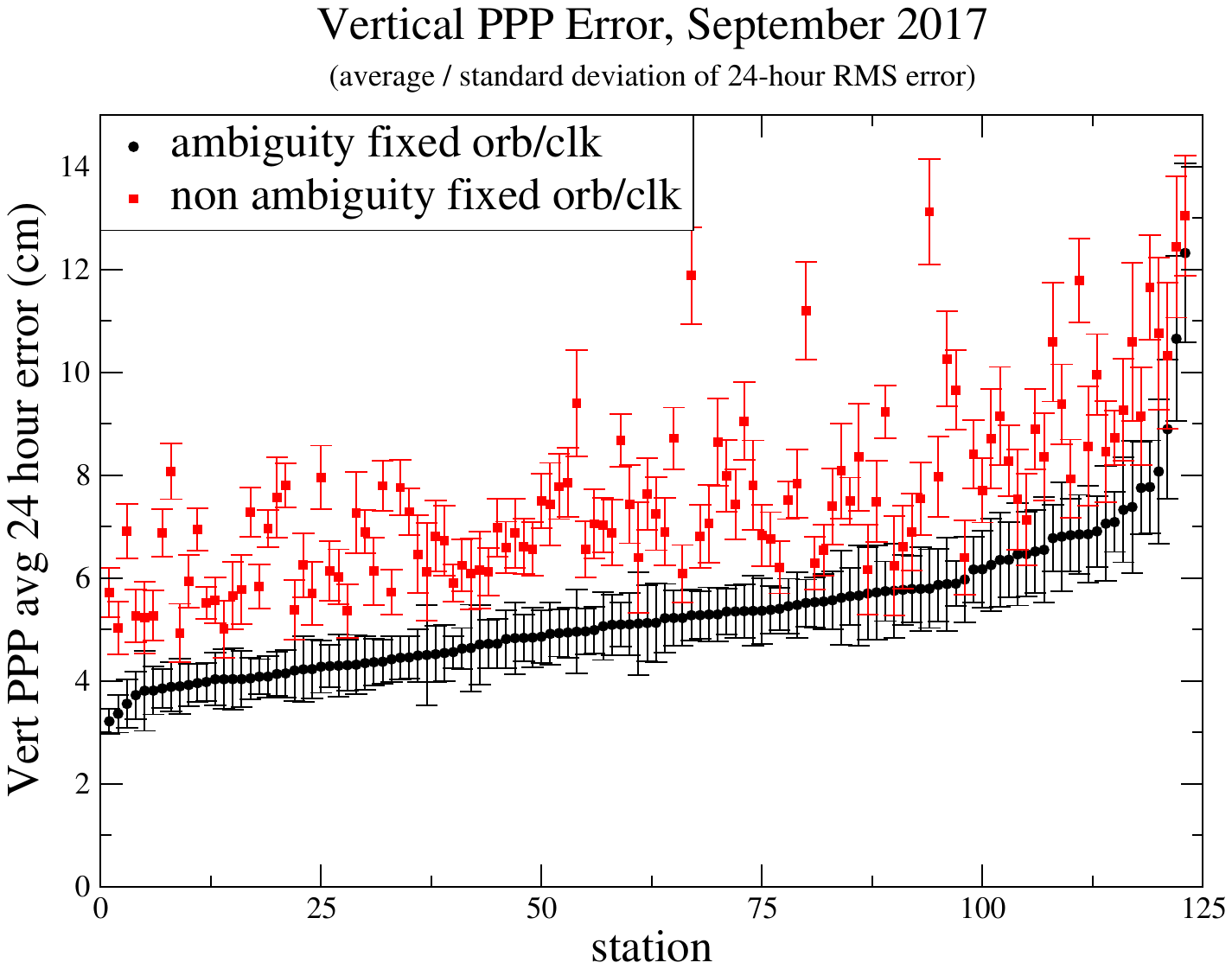} % requires the graphicx package

\caption{Horizontal (top) and vertical (bottom) real-time positioning accuracy of real-time kinematic 5-minute point-positioning of 125 GDGPS tracking sites with RTGx as decoupled partition of the GDGPS GPS orbit determination filter during September 2017. Each point represents a monthly average of daily RMS positioning error. The error bars depict the standard deviation of the daily RMS values over the entire month.  Integer phase ambiguity resolution was used to generate the black curves, and float phase ambiguity resolution was used to generate the red curves. Stations are ordered according to their integer ambiguity-resolved positioning accuracy. The reference solutions are provided by long-term static point-positioning.}

\label{fig:RTPPPAccuracy}
\end{center}
\end{figure}

The specialized requirements of real-time operations have driven the design of some unique and powerful RTGx features. One 
such feature is the ability to reconfigure a real-time running filter into partitions such that some parameters associated with 
specific satellites or specific stations, are estimated but have no influence on any other parameters. At the GDGPS System these 
``decoupled" partitions are used to accommodate unhealthy satellites in order to keep monitoring them while protecting the rest of 
the estimated constellations from the potential mismodeling of unhealthy satellites, for example due to maneuvers.  Our decoupling 
algorithm is equivalent to infinitely de-weighting the data from the associated satellite or station following a white-noise re-set of the parameters 
associated with the satellite or station. 
For earthquake monitoring the GDGPS System operates orbit determination filters with multiple decoupled partitions where the 
position of hundreds of ground sites (a site per partition) are estimated kinematically at 1 Hz \citep{larson2007improving} , but these sites are prevented from 
impacting the satellites and stations in other partitions. Only the main partition is allowed to influence other partitions. The 
incorporation of kinematic ground site positioning within an orbit determination filter, albeit in a decoupled partition, is more accurate 
than forward filtering point positioning due to the availability within the real-time filter of the full temporal correlations among all filter 
parameters. Few centimeter positioning accuracy (from high-quality, continuous tracking sites) is accomplished at 1 Hz (Fig. 
\ref{fig:RTPPPAccuracy}). Real-time time series of hundreds of ground sites are produced by the GDGPS System and made 
available for natural hazard monitoring on a global scale (via \href{https://ga.gdgps.net/}{https://ga.gdgps.net/}).

A cold start of a real-time GNSS orbit determination may take from hours to days to fully converge (i.e. to achieve the accuracy level of a 
long running continuous filter), depending on the quality of the initial states. In certain operational scenarios it is desirable to restart 
the filter with some modified configuration, while achieving instant convergence. This capability is enabled in RTGx through the 
commanded (or scheduled) saving of a ``snapshot" of the entire filter state and model state, and the ability to read this snapshot upon 
restarting to achieve a hot-start capability. By instant convergence 
we mean that if the identical data are fed to the filter after the snapshot epoch, the solution will be identical to the continuous 
running filter. This feature allows the operator to rapidly correct for errors in the past and catch back up to real-time.

Other specialized features designed to support robust real-time operations include special handling for clock references to ensure 
smooth transition from one reference clock to another, on-the-fly decoupling of satellites or sites, pairwise filter operations 
combining a low cadence orbit filter (e.g., every 60 seconds) with a high cadence (e.g. every second) clock filter for optimal throughput.

The GDGPS System produces a variety of real-time differential corrections for the GNSS broadcast ephemerides, based on the real-time orbit and clock solutions, including formats that enable real-time ambiguity resolutions such as the state-space representation standards 10403.3 of the Radio Technical Commission For Maritime Services 
 (RTCM) \href{http://www.rtcm.org/differential-global-navigation-satellite--dgnss--standards.html}{http://www.rtcm.org/differential-global-navigation-satellite--dgnss--standards.html}.

\subsection{Precise Point Positioning with GNSS}

\subsubsection{Realization of the Terrestrial Reference Frame with Precise Point Positioning}

\begin{figure}[htbp]
\begin{center}
\includegraphics[width=0.95\textwidth]{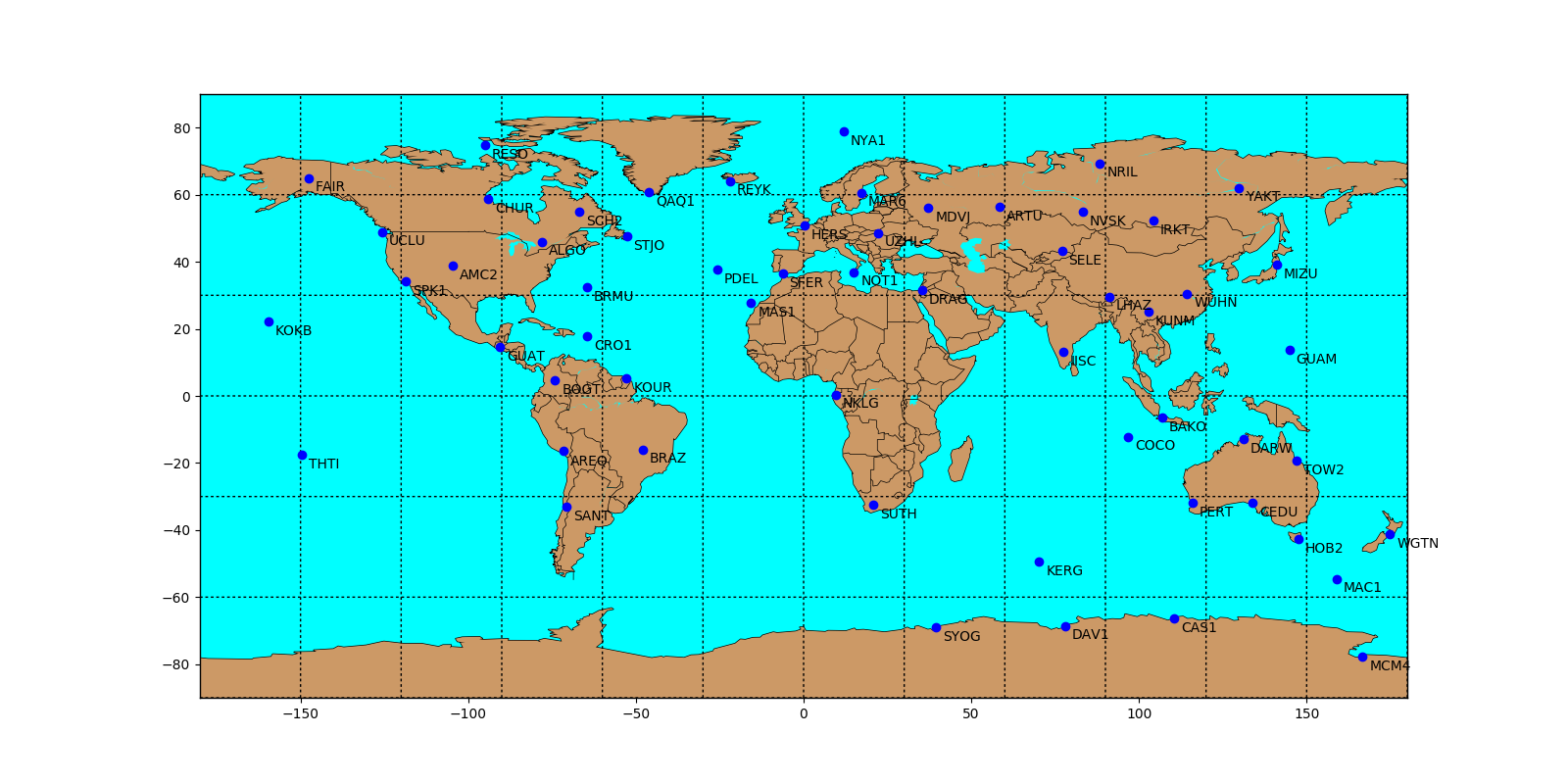} % requires the graphicx package
\caption{Geographic distribution of the 59 sites used for PPP testing.}
\label{fig:pppSites}
\end{center}
\end{figure}

In \citep{bertiger2010single}, it was demonstrated that the IGS implementation of the current ITRF frame at the time, ITRF05, could 
be realized with 24-hours of GPS data to an accuracy of 2 mm in the horizontal and 6 mm in the vertical processing 6-months of 
data from a 106 reference frame sites. Here, we do a similar experiment with 59 stations in ITRF2014.  The stations selected  
are available on at least 90\% of the days from 2002-2012, are part of IGS14 (IGS implementation of ITRF2014) \citep{rebischung2016igs}, and are selected 
to be well-distributed geographically.  Figure \ref{fig:pppSites} shows the geographic distribution of sites used for this test.

\begin{table}[htbp]
\begin{center}
\begin{tabular}{|l|r|r|r|}
\hline
PPP type & East (mm) & North (mm)  & Vertical (mm) \\
\hline 
NF no-IONEX & 1.90 & 1.98 & 6.47 \\ 
\hline 
NF & 1.89 & 2.07 & 6.49 \\
\hline 
NNR & 1.92 & 2.11 & 6.53 \\ 
\hline 
NNRTS & 1.96 & 2.11 & 6.45 \\ 
\hline 
\end{tabular} 
\caption{Median daily RMS repeatability relative to ITRF2014 solution}
\label{tab:pppRepeat}
\end{center}
\end{table}

\begin{figure}[htbp]
\begin{center}
\includegraphics[width=0.8\textwidth]{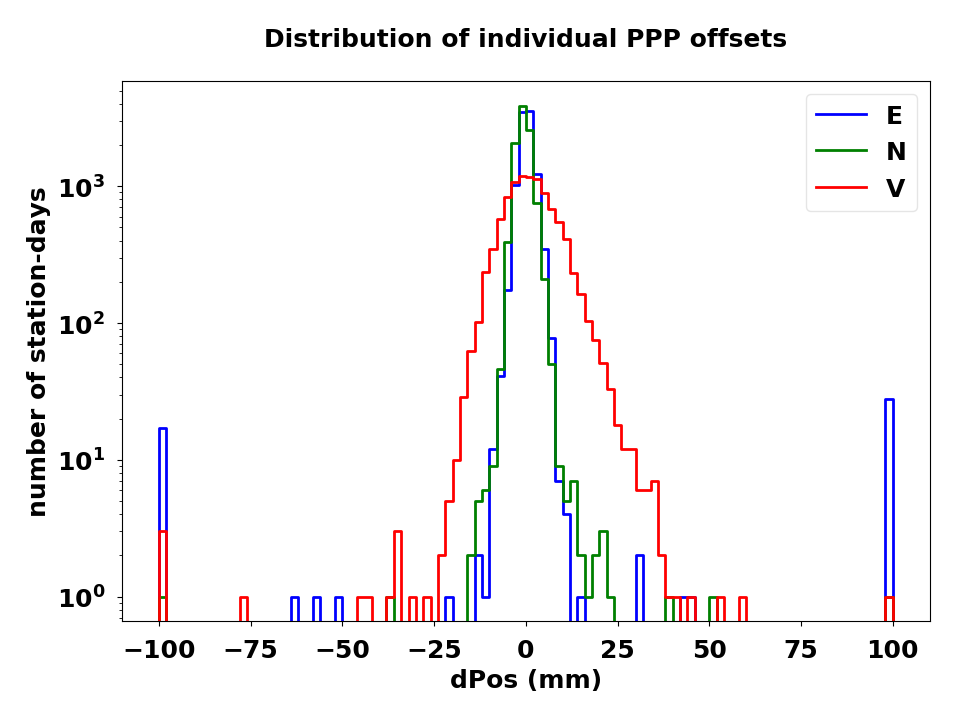} % requires the graphicx package
\caption{Histogram: East, North, and Vertical ITRF repeatability for 59 sites used for static PPP testing using NF products and IONEX corrections. Each count represents a single day of processing for a single station.}
\label{fig:pppDistro}
\end{center}
\end{figure}

First, each site was point positioned using a tree very similar to the default tree included with GipsyX-1.0.  The only substantive 
difference with the default tree is the use of Ionosphere Exchange (IONEX) files \citep{schaer1998ionex} for second-order 
ionospheric corrections (except for one no-IONEX case).  A GPT2-based \citep{bohm2015gpt2w} nominal troposphere and mapping function was used.  
Each site was positioned using JPL's repro3.0 IGS14 Final products, with separate PPP runs for the non-fiducial (NF), no-net 
rotation (NNR), and no-net translation, rotation, or scale constraint (NNRTS also referred to as ``fiducial'') products from 2008-06-01 through 2008-11-30.  In the case of
the NNR and NF products, a 7-parameter Helmert transformation is performed using the appropriate product x-file to convert the 
resulting position into the reference frame.  Positions are compared with those given in ITRF2014.   For each station, a 5-sigma 
edit is performed on total position differences, then the root-mean-square (RMS) is calculated in east, north and vertical (E, N, V) 
components.  We then report the median of each of these components in Table \ref{tab:pppRepeat}.  Figure \ref{fig:pppDistro} 
shows a histogram of the frame repeatability for all 59 stations including all outliers.
All three methods (NF, NNR, NNRTS) produce nearly identical results.
When compared to the GIPSY-OASIS results in \citep{bertiger2010single}, E and N are almost identical but the results here are a bit worse in V (6.5 vs. 6.0 mm).
We attribute the difference in vertical to station selection differences (i.e. 59 IGS14 stations vs. 106 IGS08), as vertical repeatability is highly site-dependent.
While using IONEX files does not improve repeatability in this analysis, we still recommend its use to be consistent with
the models used in the Final POD process, and because other analyses show that omitting it can introduce an offset in the z-component \citep{riesIgs2018}. In this test, the offset is an average of 0.4 mm. The ENV frame repeatability of about 2, 2, and 6 mm in Table \ref{tab:pppRepeat} is 
significantly better than the 24-hr RMS repeatability of 3, 6, and 9 mm shown in \citep{Soycan2011} using Bernese 5.0 software. We note that significant improvements 
are seen in the east component with bias fixing \citep{bertiger2010single} with the GipsyX/RTGx bias fixing methods as well as the result with PANDA software \citep{Ge2007} with bias fixing based on fractional satellite delays. In \citep{Ge2007}, they compute repeatability relative to IGS weekly solutions after removing a 7-parameter Helmert transformation and get mean RMS E, N, and V repeatability of 2.4, 2.7, and 5.3 mm with bias fixing. These are not exactly the same statistics as shown in 
Table \ref{tab:pppRepeat}. The weekly solutions will take out some vertical signal from seasonal loading not contained in the ITRF2014 frame. 

\begin{figure}[htbp]
\begin{center}
\includegraphics[width=1.0\textwidth]{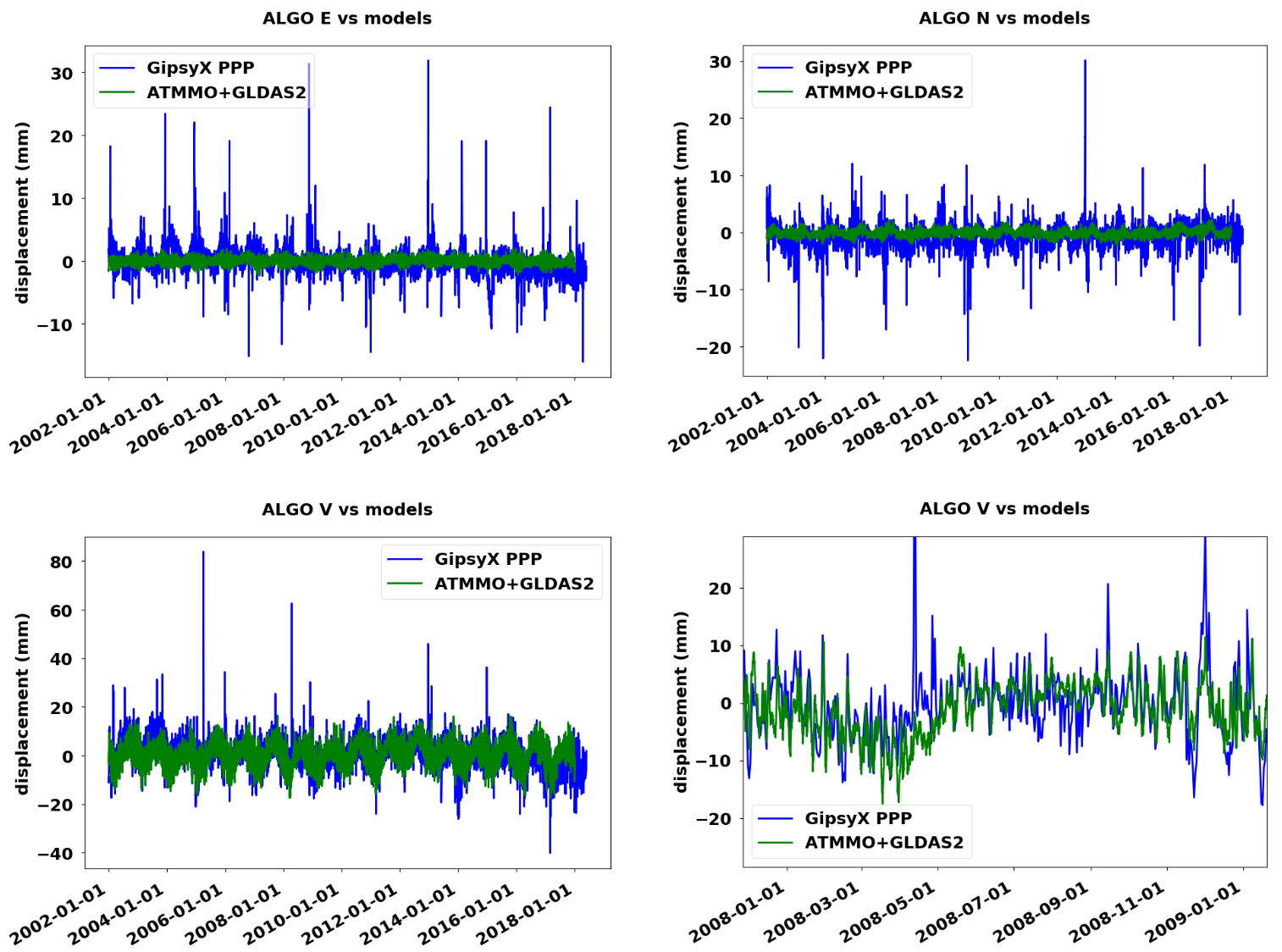} % requires the graphicx package
\caption{Daily of PPP ALGO over 13 years vs ITRF frame solution compared with displacements from \href{http://loading.u-strasbg.fr/displ\_all.php}{http://loading.u-strasbg.fr/displ\_all.php} \citep{Gegout2010} using the sum of atmospheric loading (ATMMO) and hydrological loading (GLDAS2, \citet{GLDAS}) models in E (upper-left), N (upper-right), and V (lower-left) components.  A higher temporal-resolution plot of V is provided in the lower-right.}
\label{fig:pppALGO}
\end{center}
\end{figure}

Figure \ref{fig:pppALGO} shows the seasonality of PPP residuals relative to the ITRF solution for ALGO (Algonquin Park, Canada) using IONEX corrections and NF products for 16 years years.  Most of the vertical residual can be explained by atmospheric and hydrological loading \citep{tregoning2009atmospheric}, models of which are also plotted.  The horizontal displacements also show seasonality due to other effects such as snow cover \citep{larson2013gps}.  This 
example illustrates one complexity of repeatability calculations, and how results can be biased by short time periods and/or 
inhomogenous distributions of stations.

\subsubsection{Kinematic Positioning}

% See ~wbert/gipsyX/Notes, /gdgps4/wbert/gd2ePPPframeTest2008_igs14/Kin*

For many applications the receiver is moving, and it is the accuracy of the kinematic position of the GNSS receiver that matters. 
Applications, among many, include synthetic aperture radar (SAR) from aircraft, ocean floating buoys used to determine sea 
surface height, co-seismic deformation, and ice sheet movement, see \citep{hensley2008uavsar, bonnefond2003leveling, born1994calibration, simons20112011, doake2002tide} for example. To give an idea of the expected performance from GipsyX, we 
kinematically positioned a set of 45 ITRF2014 frame stations every 5 minutes for the month of June 2008 using GPS data. Table 
\ref{tbl:kinAdjTrop} shows the RMS differences with the ITRF2014 frame positions where the position has been adjusted as a 
loose random walk, $1 m/\sqrt{s}$, while the troposphere is a bit more constrained than in the static point positioning. 
Instead of adjusting the zenith delay and tropospheric gradient parameters, only a more constrained zenith delay is adjusted to 
help break the correlation with the random walk vertical position.

\begin{table}[htp]
\caption{Kinematic positioning 45 ITRF2014 frame stations, adjusting troposphere zenith delay. For each station RMS difference relative to the station frame
position is computed for the month of June 2008. The mean and median of the station RMS values are listed.}
\begin{center}
\begin{tabular}{c c c c}
   & & RMS(mm) & \\
   & East & North & Vertical \\ \hline
%All & 6.3 & 7.2 & 20.2 \\
Mean & 6.3 & 7.2 & 20.3 \\
Median & 6.0 & 7.2 & 20.0 \\
\end{tabular}
\end{center}
\label{tbl:kinAdjTrop}
\end{table}

Some kinematic applications may have better information about the tropospheric delay. For instance, high flying aircraft typically
have a small wet delay and the dry zenith delay may be computed using a pressure sensor. To demonstrate the expected 
performance of kinematic positioning with better known troposphere, we repeated the experiment from Table \ref{tbl:kinAdjTrop}, 
but fixed the zenith delay and the gradient parameters to the values from static 24-hour position solutions. Comparing Table 
\ref{tbl:kinAdjTrop} with Table \ref{tbl:kinFixTrop}, one can see a significant and larger improvement in the vertical. 

\begin{table}[htp]
\caption{Kinematic positioning 45 ITRF2014 frame stations, tropospheric delay from static positioning, statistics as in \ref{tbl:kinAdjTrop}.}
\begin{center}
\begin{tabular}{c c c c}
   & & RMS(mm) & \\
  & East & North & Vertical \\ \hline
%All & 5.6 & 6.4 & 15.4 \\
Mean & 5.6 & 6.3 & 15.4 \\
Median & 5.4 & 6.3 & 14.9 \\
\end{tabular}
\end{center}
\label{tbl:kinFixTrop}
\end{table}

\subsection{Precise orbit and clock determination of low Earth orbiters}

Both GIPSY and RTG supported POD of satellites in Low Earth Orbit (LEO), starting with the 
launch of TOPEX/Poseidon \citep{bertiger1994gps}. As a superset of these two software, GipsyX/RTGx supports 
precise post-processing, reduced-dynamic orbit determination, as well as embedded real-time POD. For post-processed LEO POD using GPS data, the GipsyX/RTGx provides identical capability to GIPSY, but with a simpler user interface along with additional capability for the newer GNSS constellations, DORIS, and SLR. Tests were performed with GRACE \citep{tapley2004grace}, the follow-on 
GRACE mission (GRACE-FO), and Jason-2 \citep{lambin2010ostm, bertiger2010sub, cerri2010precision}. Since the GPS determined 
orbits for Jason-2 are sub-centimeter accurate \citep{bertiger2010sub}, we can use those orbits to measure our accuracy for Jason-2 orbits  
determined with other data types. 
Mixing GNSS, SLR, and DORIS data is also possible in GipsyX/RTGx, but has not 
had extensive testing yet. In the subsections below, we detail the expected accuracy and software capabilities for LEO POD using GPS, SLR, and 
DORIS data.

\subsubsection{SLR Orbit determination}
GipsyX has many capabilities for SLR processing, including producing residuals to orbits determined from other techniques and independent orbit determination.
For one experiment, We used 30-hour arcs of SLR data from 15 well-distributed, low-bias SLR sites, centered on noon of each day in 2015 to determine Jason 2's orbit with a median radial RMS difference to GPS-determined orbits of 1.9 cm.
We also validated our SLR capabilities by performing orbit determination on the LAGEOS 1 and 2 satellites using 7-day arcs for all of 2017.  We then compared our orbits to the ILRSA combined orbits.  Our LAGEOS 1 and 2 orbits have an overall RMS difference of 5 mm in radial, 16 mm in cross-track and 18 mm in along-track to LAGEOS, which is comparable in magnitude to the inter-center differences measured by the ILRSA combination to other centers over the course of the year (averages of 5 mm, 24 mm, and 25 mm in the weekly ilrsa weekly sum files across all centers for both LAGEOS satellites).

\subsubsection{GPS --- GRACE, GRACE-FO POD and time synchronization}

GipsyX/RTGx is the operational software used for GRACE-FO to determine the orbit and clock of the GRACE-FO spacecraft. 

\begin{figure}[htbp]
\begin{center}
\includegraphics[width=0.8\textwidth]{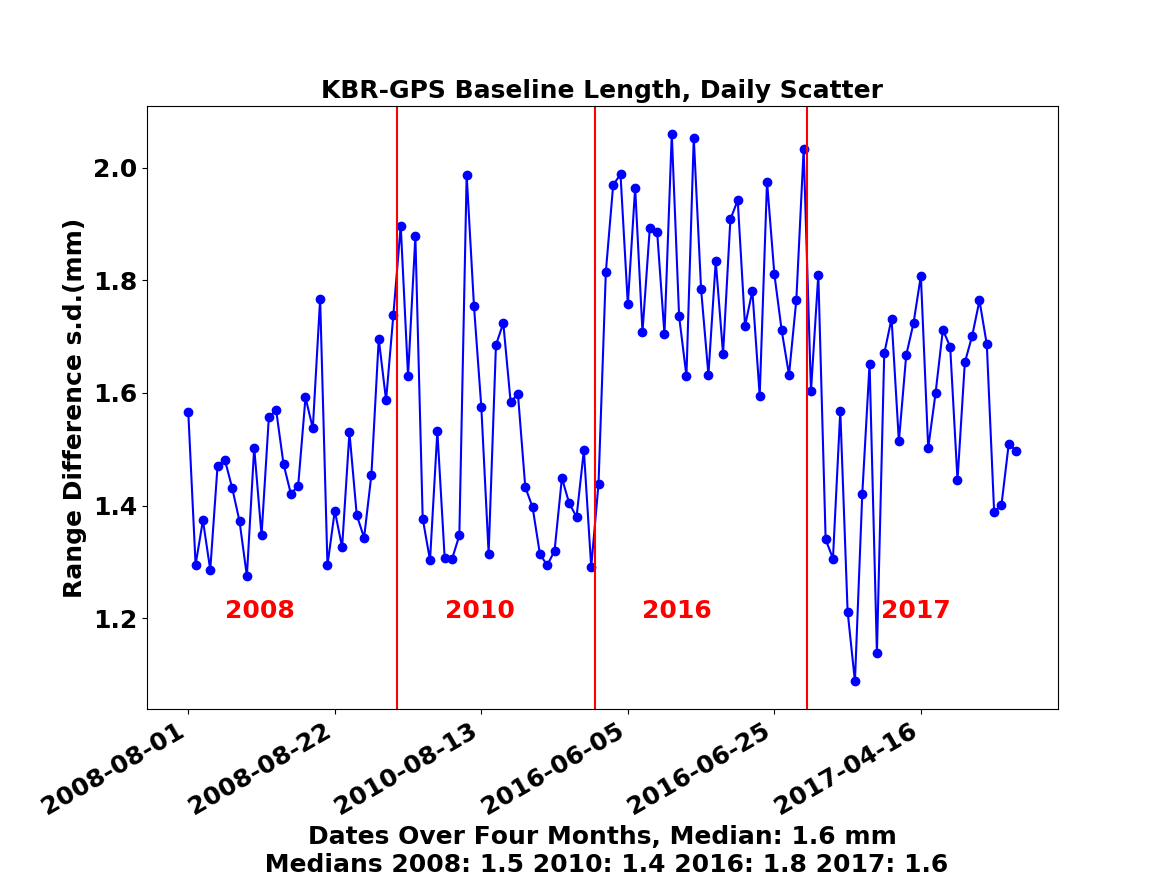} % requires the graphicx package
\caption{GRACE KBR-GPS Range, Daily GPS POD Baseline Accuracy, Four 1-Month Samples: Aug. 2008, Aug. 2010, June 2016, April 2017}
\label{fig:KBR-GPS4Months}
\end{center}
\end{figure}

GRACE and GRACE-FO are among the best low Earth orbiting satellites to test POD with GPS since they have a 
measurement of the inter-satellite range (up to a bias) that is good to the micron-level. This is the dual-one way range measurement 
made at K and Ka band, K-Band Range (KBR) \citep{dunn2003instrument, thomas1999}. It is the KBR's precision along with an accelerometer measuring the non-gravitational forces that 
allows the precise recovery of the Earth's time varying gravity field. Although GipsyX/RTGx, can process KBR range data and use accelerometer data, neither is used in the operational POD for GRACE or GRACE-FO discussed here. Figure \ref{fig:KBR-GPS4Months} shows the daily 
standard deviation of the difference of the range determined by GPS POD and the KBR (clocks aligned with the 
independent operational code). The processing used GPS data sampled every 10-sec, antenna calibrations based on one 
month of data, August 2010, reduced dynamics selected to perform well over the full time period, and bias fixing. The 
average daily baseline standard deviation was 1.6 mm. This is significantly better than the 10 mm found in Table 8 of \citep{Kang2006} using the Center for Space Research's
MSODP software which does not include bias fixing. It improves over the results in \citep{bertiger2010single} mostly due to the use of higher rate data. \citep{Kroes2005} and \citep{Jaggi2007} show improved baseline determination of a little under a mm if biases are fixed between the two spacecraft with a simultaneous POD of both spacecraft. In this processing, both spacecraft are processed independently.

\begin{figure}[htbp]
\begin{center}
\includegraphics[width=0.8\textwidth]{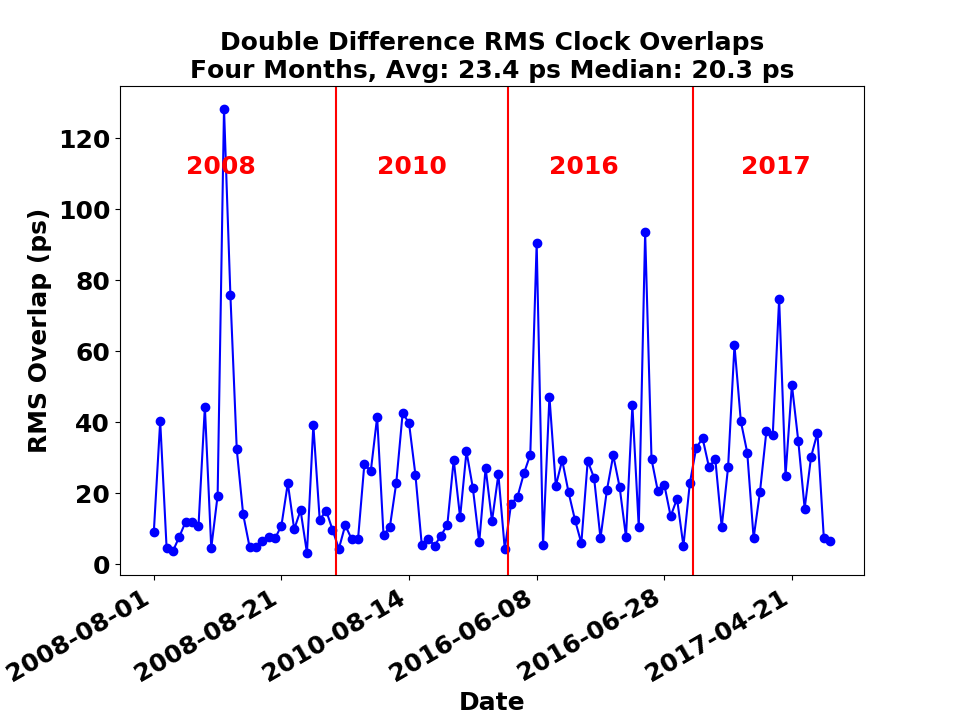} % requires the graphicx package
\caption{RMS Double Difference GRACE Clock Overlaps, Four 1-Month Samples: Aug. 2008, Aug. 2010, June 2016, April 2017}
\label{fig:DDclk}
\end{center}
\end{figure}

Although the KBR measurement is not strongly dependent on the GPS POD/Clock solution for GRACE at the 100 
micron level, it is necessary to have the error of the relative clocks between the two GRACE spacecraft less than 160 ps, 1-sigma, 
after removing a bias (see \citep{thomas1999} eq. 3.19) to have the errors in the KBR range due to the clock alignment less than 0.5 microns. 
Figure \ref{fig:DDclk} shows a measure of the precision of the relative clock between the two GRACE spacecraft. 
Since we are solving for the GRACE clock as a white noise process, relative to a fixed GPS reference with 30-hours of 
data centered on noon of each day, there is a six hour overlap from one day to the next. If our reference clock were the 
same, and our solution were perfect, the difference of the clock solutions between these two days would be zero. In 
eq. \ref{eq:DDclk}, the first term is the difference of the two GRACEA clocks, the second GRACEB. Taking the 
difference of the A and B differences removes any reference clock, thus the RMS of this quantity over the central 5
hours of the 6-hours is a good measure of the relative clock precision. Fig. \ref{fig:DDclk} shows that we are 
significantly under the 160 ps requirement with a median value of 20.3 ps.

\begin{equation}
\label{eq:DDclk}
dd_{clk} = (A_{clk}^{day1} - A_{clk}^{day2}) - (B_{clk}^{day1} - B_{clk}^{day2})
\end{equation}

\subsubsection{Doppler System POD}

In addition to GNSS phase and range data, GipsyX/RTGx can process almost any radio-metric phase and range data and 
is easily extended to data types that are linear combinations of radio-metric range/phase data; for instance, Satellite 
Laser Ranging or {Doppler Orbitography and Radiopositioning Integrated by Satellite} (DORIS) \citep{willis2010international}. Current DORIS data is given in a RINEX like format \citep{auriol2010doris} with dual-frequency phase 
measurements and range measurements similar to GNSS. Traditionally this is processed as Doppler

\begin{equation}
\label{eq:Dop}
   D = \frac{\phi(\tau_1) - \phi(\tau_2)}{\tau_1 - \tau_2}
\end{equation}
where $\phi(\tau)$ is the measured phase on an orbiter from a ground transmitter at it's local time $\tau$. Prior 
to  June 20, 2008 the Doppler measurements were given directly with corrections for errors in the satellite clock and phase 
center offsets. The count time, $\tau_1 - \tau_2$, is 10 seconds of local satellite time in eq. \ref{eq:Dop}. When 
forming DORIS Doppler from the phase on the RINEX DORIS data, one must model the satellite receiver clock errors 
for the computed Doppler model, $C$, 

\begin{equation}
\label{eq:CompDop} 
\begin{aligned}
 C  &= \frac{\psi(\tau_1) - \psi(\tau_2) + c(E(\tau_1) - E(\tau_2))}{t_1 - t_2 + E(\tau_1) - E(\tau_2)} \\[1em]
      &= \frac{  \frac{\psi(\tau_1) - \psi(\tau_2)}{t_1-t_2} + \frac{c(E(\tau_1) - E(\tau_2))}{t_1-t_2}  }{1 + \frac{E(\tau_1) - E(\tau_2)}{t_1-t_2}}
\end{aligned}
\end{equation}
The phase measurement, with units of length, is written as $\phi(\tau) = \psi(\tau) + cE(\tau)$, where $E(\tau)$ is 
the difference between coordinate time ($t$) and the local time of the receiver in seconds and $c$ is the speed 
of light. For the DORIS system, the receiver is driven by a stable clock (Allan deviation on the order of $10^{-13}$ at a day) and the 
rate of the clock errors including relativistic effects, $\frac{E(\tau_1) - E(\tau_2)}{t_1-t_2}$ is small, typically on the 
order of $10^{-7}$ s/s, we can expand eq. \ref{eq:CompDop}

\begin{equation}
\label{eq:CompExpand} 
\begin{aligned}
C \approx &\frac{\psi(\tau_1) - \psi(\tau_2)}{t_1-t_2} + c\frac{E(\tau_1) - E(\tau_2)}{t_1-t_2} \\[3pt]
    &- \frac{E(\tau_1) - E(\tau_2)}{t_1-t_2} \frac{\psi(\tau_1) - \psi(\tau_2)}{t_1-t_2}  \\[3pt]
    &- c \left(\frac{E(\tau_1) - E(\tau_2)}{t_1-t_2}\right)^2
\end{aligned}
\end{equation}
The computed measurement model in eq. \ref{eq:CompExpand} is very close to the difference of phase 
measurements and fairly easy to code in terms of the basic phase measurement used for GNSS. This model also 
may be used for the legacy DORIS doppler data since the clock error terms are essentially zero, due to pre-calibration. 
On the DORIS RINEX file, the nominal receiver clock values are given as a time series derived from the range data. A 
quadratic fit over time periods on the order of a day maybe used for $E(\tau)$. 

\begin{figure}[htbp]
\begin{center}
\includegraphics[width=0.8\textwidth]{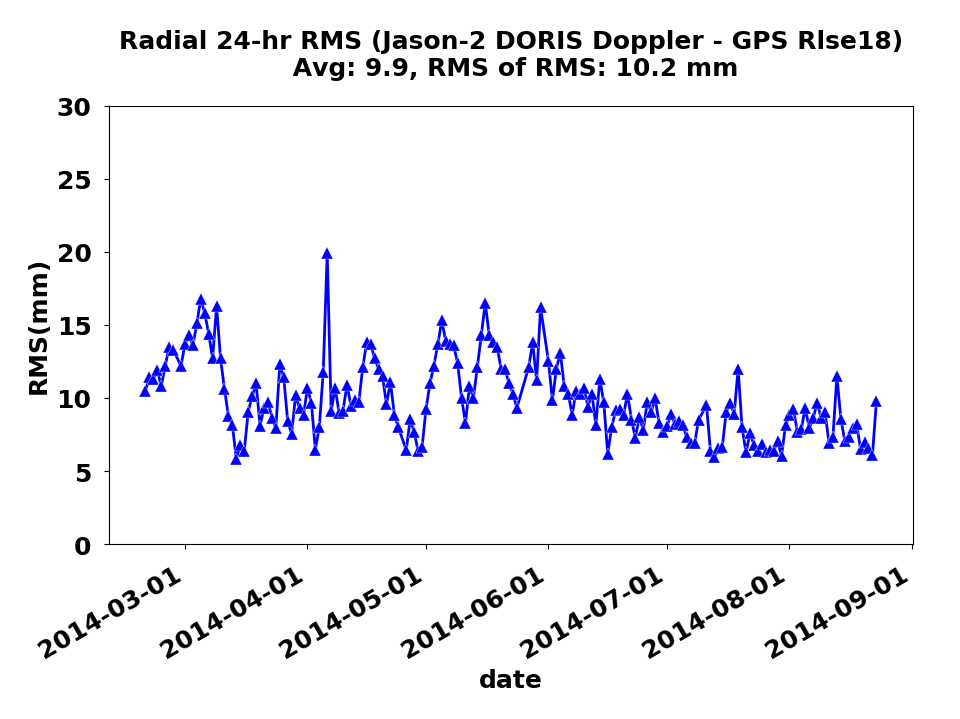} % requires the graphicx package
\caption{RMS Radial DORIS Doppler radial orbit errors relative to GPS JPL RLSE18}
\label{fig:DORISradialErr}
\end{center}
\end{figure}

 We have tested this with data from the Jason-2 spacecraft \citep{couhert2015towards} using station coordinates from the latest  DPOD (DORIS Precise Orbit Determination) solution, DPOD2014,  aligned on ITRF2014 \citep{moreaux2019dpod2014}. To take into account the effect of the South Atlantic Anomaly 
(SAA) on the onboard oscillator \citep{willis2016jason}, we did not use any correction model but removed a few stations in the South 
America region. In Fig. \ref{fig:DORISradialErr}, we compare the Jason-2 radial orbit position determined with the DORIS RINEX 
data from Feb. 14, 2014 through August 23, 2014 with a GPS-determined orbit whose radial accuracy is about 5 mm 
\citep{Ostst2018} using ITRF2014 coordinates for tracking stations and typical RMS differences with the GipsyX/RTGx DORIS determined orbit are at the 9.9 mm level. Errors in radial position go directly into the mission's prime 
objective to measure sea surface height and are the most important metric for ocean altimetry missions.

To better evaluate the SAA effect on Jason2, we will soon investigate which stations are the most affected and use the new available correction model from \citep{belli2018long}, based on T2L2 measurements. In the future, we plan to analyze the Sentinel-3A DORIS RINEX data. The fact that satellite clock for Sentinel-3A is the same for the GPS and the DORIS on-board instruments should allow us to compare the GPS clock solution and the DORIS clock.  The GPS solution for the on-board clock could be used to properly correct the SAA effects for this satellite, \citep{jalabert2018analysis}.

\subsection{Multi-Technique Geodesy}

As an example of multi-technique use, we processed GPS and SLR data simultaneously using GPS and SLR data from low Earth orbiters
to tie the systems together. For the development of reference frames, the ties are traditionally made through local ground surveys to measure the vector between the GPS and SLR instruments. With only three days of 
data, and no ground survey ties, we show agreement at the cm level.

\begin{figure}[htbp]
\begin{center}
\includegraphics[width=0.8\textwidth]{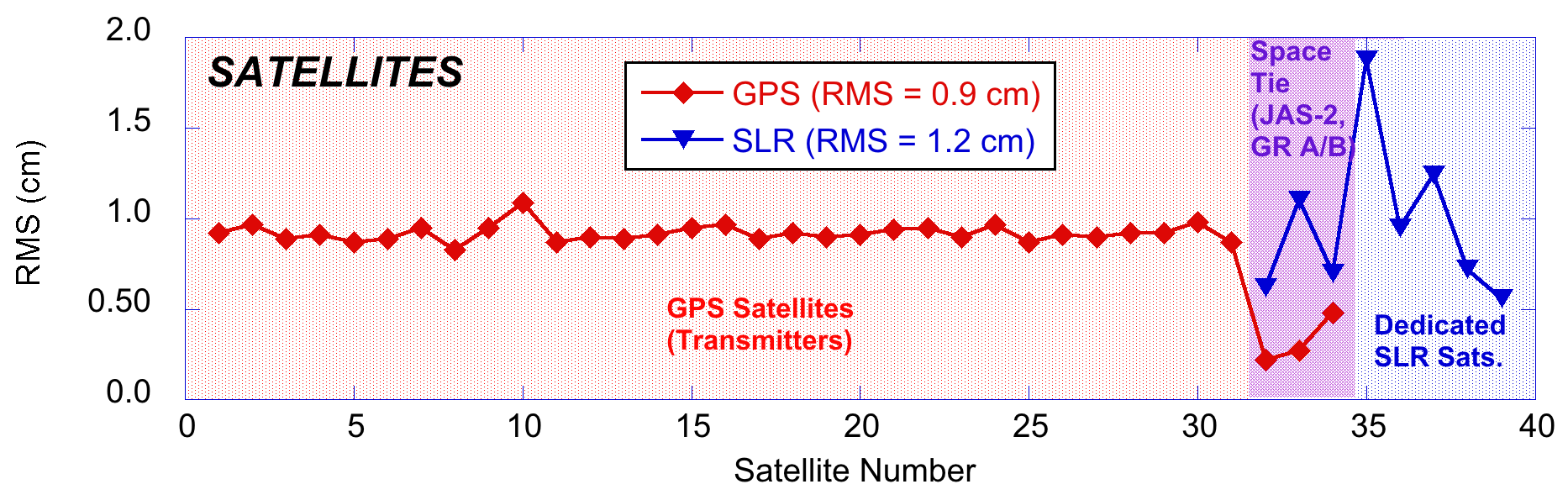} % requires the graphicx package
\caption{GPS dual-freq. phase residuals and SLR residuals combining SLR and GPS Data; 
Jason-2, GRACE A, B $=$ Sat. Number 32, 33, 34; LAGEOS 1, 2, Starlette, Stella,  Ajisai $=$ 35, 36, 37, 38, 39. }
\label{fig:SLR_GPS_Res}
\end{center}
\end{figure}

Depicted in Figure \ref{fig:SLR_GPS_Res} are summaries (by satellite) of tracking data residuals from a multi-day GipsyX/RTGx solution 
in which SLR and GPS are combined at the observation level. Represented in the solution are 31 GPS satellites, 5 dedicated SLR 
targets in space (LAGEOS 1 and 2, Starlette, Stella and Ajisai), and 3 low-Earth orbiters (LEO) with both GPS receivers and corner 
reflectors (Jason-2 and GRACE A and B). The latter three LEO missions provide a means of accurately linking the SLR and GPS 
systems without the benefit of ground survey ties. The uniform and precise (cm-level) tracking residuals across both data types 
(SLR and GPS) testify to the coherence of these diverse space-geodetic observations (radiometric and optical) in this grand 
GipsyX/RTGx network solution.

The solution also features 45 and 18 GPS and SLR ground stations respectively. To compensate for know systematic SLR errors,
range biases (per satellite) were estimated for a minority (40\%) of the stations.  Satellite-specific estimates of the optical range
variations as a function of the line-of-sight could further improve the results, but were not applied in this case. 
The positions of all satellites and stations were 
estimated simultaneously in the GipsyX/RTGx run (ref.), along with Earth orientation. Using this `fiducial-free' strategy (ref.), the network 
of satellites and stations will be subject to rotation, but the scale, origin, and relative positions will be well determined. Two of the 
SLR stations in the run (Matera, Italy and Tahiti, French Polynesia) also have participating GPS stations linked by ground survey ties. With only 3 days of 
tracking data, the GipsyX/RTGx solution reproduced these independent ground ties to ~1 cm (3D). The tie between SLR and GPS in the 
GipsyX/RTGx solution comes solely from the space ties on Jason-1 and the GRACE tandem, testifying not only to the benefits of 
combining data at the observation level but also to the fidelity of the GipsyX/RTGx modeling.

\subsection{Building Terrestrial Reference Frames and Fitting Ground Site Time Series}

GipsyX includes station coordinate processing software which can combine solutions from several techniques to build reference 
frames, simulate future performance, compute transformation parameters between reference frames, and fit individual time series to estimate 
positions, velocities, 
seasonals, and possible discontinuities in the linear and periodic fit, which we refer to as breaks.  Each tool has command line help available.  On-line 
training provides detailed instructions and example 
data sets for the most common use cases.

\begin{table}[htp]
\begin{center}
\caption{Comparison of GipsyX frame with ITRF2014 on January 1, 2005.  Three translations, one scale, and three rotations are given in mm while the corresponding rates are given in mm/yr.  Rotations and scale are multiplied by the Earth radius to get mm at the surface.}
\label{tab:Frame1}
\begin{tabular}{l|r|r|r|r|r|r|r|l}
\textbf{Parameter} & \textbf{TX} & \textbf{TY} & \textbf{TZ} & \textbf{S} & \textbf{RX} & \textbf{RY} & \textbf{RZ} & \textbf{Unit}\\
\hline
Offset         &  0.508   & -0.192   & -0.144   &  2.903   &  0.543   & -0.053   & -0.354 & mm \\
Offset s.d.   &  0.102   &  0.095   &  0.200   &  0.521   &  0.117   &  0.141   &  0.204 & mm \\
Rate           & -0.184   &  0.044   &  0.498   &  0.153   &  0.001   &  0.076   &  0.156 & mm/yr \\
Rate s.d.     &  0.013   &  0.012   &  0.027   &  0.036   &  0.018   &  0.021   &  0.026 & mm/yr \\
\end{tabular}
\end{center}
\end{table}

\begin{table}[htp]
\begin{center}
\caption{Comparison of GipsyX frame with DTRF2014 on January 1, 2005.  Three translations, one scale, and three rotations are given in mm while the corresponding rates are given in mm/yr.  Rotations and scale are multiplied by the Earth radius to get mm at the surface.}
\label{tab:Frame2}
\begin{tabular}{l|r|r|r|r|r|r|r|l}
\textbf{Parameter} & \textbf{TX} & \textbf{TY} & \textbf{TZ} & \textbf{S} & \textbf{RX} & \textbf{RY} & \textbf{RZ} & \textbf{Unit}\\
\hline
Offset         &  0.434   & -0.032   & -0.861    & 0.049    & 0.693    & 1.221   & -1.023 &mm \\
Offset s.d.   &  0.078    & 0.077   &  0.174    & 0.079    & 0.048    & 0.067   &  0.079 &mm \\
Rate           & -0.112    & 0.077    & 0.629    & 0.057   & -0.098   & -0.121    & 0.329 &mm/yr \\
Rate s.d.     &  0.009    & 0.010    & 0.021    & 0.012    & 0.010    & 0.012    & 0.014 & mm/yr \\
\end{tabular}
\end{center}
\end{table}

A terrestrial reference frame is typically defined by a set of reference positions with a model of time 
evolution, the simplest example being a table of positions and velocities. 
Reference frames are built by first combining daily or weekly solutions for each individual technique and then combining all 
techniques with ties into a single frame.  A terrestrial reference frame built using GipsyX with inputs from all four geodetic techniques and ties is
compared to ITRF2014 \citep{altamimi2016itrf2014} in Table \ref{tab:Frame1} and DTRF2014 \citep{seitz2016new} in Table \ref{tab:Frame2}.
See \href{file:ElectronicS3_Frame.pdf}{electronic supplement 3} for details.

GipsyX includes a network simulation tool which can generate output files at a given frequency for a specified time span, for example daily files spanning one year.
The user provides a model consisting of positions, velocities, breaks, and/or seasonal terms.
Various forms of white noise can then be injected.  Coordinate noise can be added which is independent from site to site.
Geocenter, scale, and/or rotational noise can be injected which is correlated and impacts all sites.
Simulated solutions can be processed just like real ones to test software or predict performance of future networks.

\begin{figure}[htbp]
\begin{center}
\includegraphics[width=0.8\textwidth]{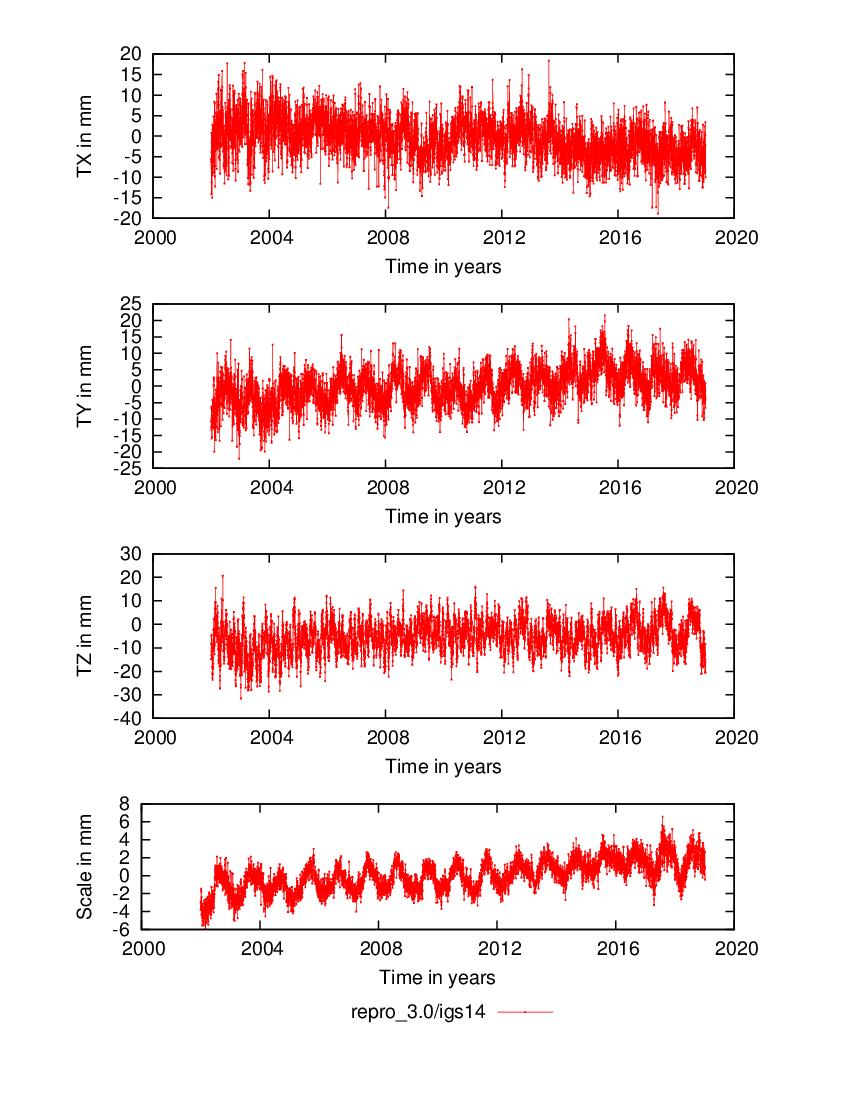} % requires the graphicx package
\caption{X-file geocenter and scale parameters relative to IGS14.}
\label{fig:Xfile}
\end{center}
\end{figure}

The transformation between two reference frames at a particular time can be described by three translations, three rotations, and 
one scale.
Daily transformation parameters known as x-files are computed using an input reference frame to predict coordinates at a particular epoch and then using those predictions 
to estimate transformation parameters to GNSS positions observed at that epoch.  
Daily x-files are publicly available as described in section \ref{OpsProducts}.
A plot of x-file parameters between IGS14 and GipsyX free-network solutions is shown in Figure \ref{fig:Xfile} .  The TX 
parameter has a mean value of -0.5 mm and a standard deviation of 4.7 mm.  The TY parameter has a mean value of -0.7 mm 
and a standard deviation of 5.6 mm.  The TZ parameter has a mean value of -5.3 mm and a standard deviation of 6.8 mm.  The 
Scale parameter has mean value of 0.0 mm and a standard deviation of 1.7 mm.  These daily corrections indicate how far GNSS 
only solutions are from a standard reference frame such as IGS14.  The GNSS only geocenter values have approached 
IGS14 more closely over time as modeling and data quality have improved and recent research has shown that adding LEO satellites such as 
GRACE and Jason allow GNSS to be competitive with SLR for both linear and annual geocenter estimation 
\citep{haines2015realizing}.

Time series for an individual site in a particular reference frame can be fit to estimate positions, velocities, breaks, and seasonals.  GipsyX 
includes tools for automatic break detection and outlier removal.  Break detection carries out an exhaustive search based on all 
possible break locations.  An F-test is carried out for the location which minimizes chi squared per degree of freedom.  If the F-test 
is passed that break is kept and the search continues at all remaining possible break locations.  The search stops when no 
remaining break location satisfies the F-test.  Outlier removal is based on a maximum value for either the sigma of each point or 
the residual/sigma ratio for each point.  Estimation is carried out using the same time variable square root information filter used 
for reference frame combinations.

\begin{figure}[htbp]
\begin{center}
\includegraphics[width=0.8\textwidth]{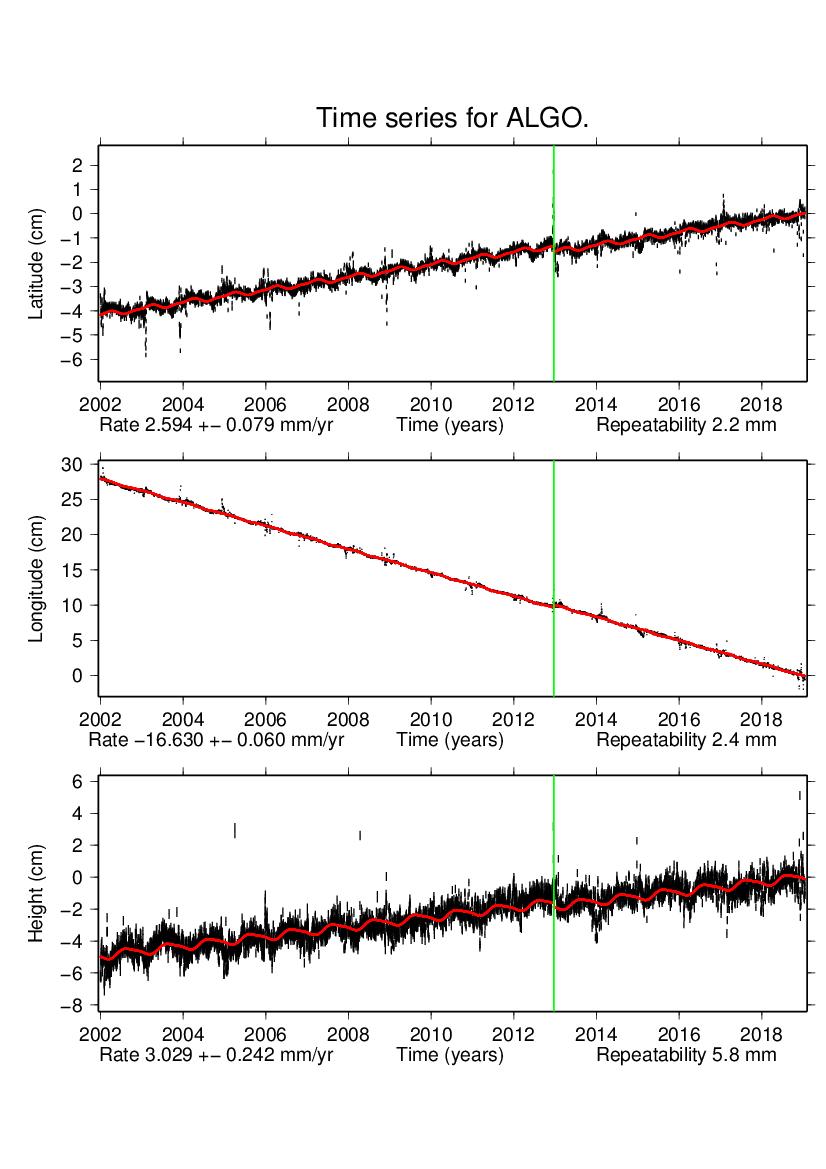} % requires the graphicx package
\caption{Time series and model fit for ALGO, Algonquin Park, Canada.}
\label{fig:ALGO}
\end{center}
\end{figure}

Fitting tools have been optimized to compute large velocity fields quickly.  Processing can be done in parallel because point 
positions at different sites are independent.  Given 2000 global sites with daily measurements spanning 20 years; break detection, 
outlier removal, and parameter estimation can be carried out in about 2 hours on a cluster containing
thirty two Intel(R) Xeon(R) X5450 @ 3.00GHz processors with eight core nodes each.
GipsyX is used to update parameter estimates and edited time series for thousands of globally 
distributed GNSS sites every week and and they are made available via our web service \url{https://sideshow.jpl.nasa.gov/post/
series.html}.  Results for the site ALGO are shown in Figure  \ref{fig:ALGO}.  The plot shows daily measurements with error bars 
in black,  the model fit in red, and break times in green.  Velocity estimates with error bars are provided along with residual 
repeatability/WRMS values for each component.

\section{Summary}

The new GipsyX/RTGx software is a robust and powerful tool for geodetic data analysis and simulations, incorporating decades of expertise and lessons learned from the design and operations of previous software generations, and their applications to the most challenging positioning, navigation, timing, and science applications.  GipsyX/RTGx now underlie all GNSS orbit determination operations at JPL, and with hundreds of academic and research licenses, powers geodetic analyses and science operations across the globe. 

\section{Acknowledgements}
We are grateful to the following individuals for contributions to GipsyX/RTGx: Jason Gross, Miquel Garcia Fernandez, Jan Weiss. 

The research was carried out at the Jet Propulsion Laboratory, California Institute of Technology, under a contract with the National Aeronautics and Space Administration. 

We acknowledge support from the U.S Air Force GPS OCX program, from the JPL Global Differential GPS (GDGPS) System, and from NASA’s Space Geodesy Program.

Part of this work was performed supported by the \textit{Centre National d'Etudes Spatiales} (CNES) and is based on DORIS satellite observations.

\textsuperscript{\textcopyright}2018 California Institute of Technology. Government sponsorship acknowledged.

%\section*{References}
\newpage
\bibliography{GipsXjasr2019}

\end{document}